\documentclass[a4paper,12pt]{article}
\pdfoutput=1 % if your are submitting a pdflatex (i.e. if you have
\usepackage{amsmath}
\usepackage{amssymb}
\usepackage{epsfig}
\usepackage{graphicx}
\usepackage{cite}
\usepackage{color}

\allowdisplaybreaks[4] % for page break

\newcommand{\be}{\begin{equation}}
\newcommand{\ee}{\end{equation}}
\newcommand{\beq}{\begin{eqnarray}}
\newcommand{\eeq}{\end{eqnarray}}

\def\nue{\mathrel{{\nu_e}}}
\def\numu{\mathrel{{\nu_\mu}}}

\def\barnue{\mathrel{{\bar \nu}_e}}
\def\barnumu{\mathrel{{\bar \nu}_\mu}}

\def \lta {\mathrel{\vcenter{\hbox{$<$}\nointerlineskip\hbox{$\sim$}}}}
\def \gta {\mathrel{\vcenter{\hbox{$>$}\nointerlineskip\hbox{$\sim$}}}}

\def\t13{\mathrel{{\theta_{13}}}}
\def\y12{\mathrel{{\tan^2 \theta_{12}}}}
\def\c2{\mathrel{{\chi^2 }}}
\def\msun{\mathrel{{M_\odot }}}
\def\rsun{\mathrel{{R_\odot }}}

\def \oss		{oscillations}

\newcommand{\n}{neutrino}
\newcommand{\ns}{neutrinos}
\newcommand{\ic}{IceCube}
\newcommand{\td}{TDE}
\newcommand{\bh}{SMBH}
\newcommand{\tds}{TDEs}
\newcommand{\sft}{Swift J1644+57}
\newcommand{\sftt}{Swift J2058.4+0516}
\newcommand{\sfttt}{Swift J1112.2-8238}

\newcommand{\bi}{\begin{itemize}}
\newcommand{\ei}{\end{itemize}}

\newcommand{\ie}{{\it i.e.}}

\newcommand{\eg}{{\it e.g.}}

\newcommand{\cf}{{\it cf.}}

\newcommand{\etc}{{\it etc.}}
\newcommand{\eq}{Eq.}
\newcommand{\eqs}{Eqs.}

\newcommand{\fig}{Fig.}
\newcommand{\Fig}{Fig.}

\newcommand{\Ref}{Ref.}
\newcommand{\Refs}{Refs.}
\newcommand{\Sec}{Sec.}

\newcommand{\App}{App.}

\newcommand{\Tab}{Table}

\newcommand{\mmin}{M_{\mathrm{min}}}
\newcommand{\mmax}{M_{\mathrm{max}}}

\newcommand{\equ}[1]{\eq~(\ref{equ:#1})}
\newcommand{\figu}[1]{\fig~\ref{fig:#1}}

\begin{document}

\begin{titlepage}

\renewcommand{\thefootnote}{\alph{footnote}}

% \vspace*{-3.cm}
% \begin{flushright}
% 
% \end{flushright}

%\vspace*{0.5cm}

\renewcommand{\thefootnote}{\fnsymbol{footnote}}

{\begin{center}
{\large\bf
High Energy Neutrinos from the  Tidal Disruption of  Stars
} 

\end{center}}

\renewcommand{\thefootnote}{\alph{footnote}}

\vspace*{.3cm}
{\begin{center} {\large{\sc  
                Cecilia~Lunardini\footnote[1]{\makebox[1.cm]{Email:}
                cecilia.lunardini@asu.edu} and
                Walter~Winter\footnote[2]{\makebox[1.cm]{Email:}
                walter.winter@desy.de}
                }}
\end{center}}
\vspace*{0cm}
{\it
\begin{center}

\footnotemark[1]
          Department of Physics, Arizona State University, \\ 450 E. Tyler Mall, Tempe, AZ 85287-1504 USA
\\[0.3cm]
\footnotemark[2]
          Deutsches Elektronen-Synchrotron (DESY), Platanenallee 6, \\ D-15738 Zeuthen, Germany

\end{center}}

\vspace*{1cm}

\begin{center}
{\Large \today}
\end{center}

{\Large \bf
\begin{center} Abstract \end{center}  }

 We study the production of high energy neutrinos in jets from the tidal disruption of stars by supermassive black holes. The diffuse neutrino flux expected from these tidal disruption events (TDEs) is calculated both analytically and numerically, taking account the dependence of the rate of  TDEs on the redshift and black hole mass.
We find that $\sim 10\%$ of the observed diffuse flux at IceCube at an energy of about 1~PeV can come from TDEs if the characteristics of known jetted tidal disruption events are assumed to apply to the whole population of these sources. 
If, however, plausible scalings of the jet Lorentz factor or variability timescale with the black hole mass are taken into account, the contribution of the lowest mass black holes to the neutrino flux is enhanced.  In this case, TDEs can account for most of the neutrino flux detected at IceCube, describing both the neutrino flux normalization and spectral shape with moderate baryonic loadings.
 While the uncertainties on our assumptions are large, a possible signature of TDEs as the origin of the IceCube signal is the transition of the flux flavor composition from a pion beam to a muon damped source at the highest energies, which will also result in a suppression of Glashow resonance events.  
\\[1cm]
%@arxiver{scaledplot.pdf}

\vspace*{.5cm}

\end{titlepage}

\newpage

\section{Introduction} 
\label{sec:intro}

It is an established fact that supermassive black holes (\bh) inhabit the center of most or all galaxies. The physics of these objects is still mysterious in many ways, and can be studied by observing the effects that the enormous gravitational field of a \bh\  produces on the surrounding gas and stellar matter. 
 
A particularly dramatic effect is a Tidal Disruption Event (\td), the phenomenon in which a star passing within a critical distance from the \bh\ is torn apart by its extremely strong tidal force. 
The accretion of the disrupted stellar matter on the \bh\ can generate observable flares of radiation in the thermal, UV and X-rays that might last for days, months, or even years \cite{Hills75,Rees:1988bf,Lacy82,Phinney89}.   These flares have the potential to reveal important information on the innermost stellar population of a galaxy, and on the physics of  \bh.  \tds\ are especially valuable as probes of \bh\ that are normally quiet -- as opposed to the Active Galactic Nuclei -- and therefore more difficult to study.  
In a recent catalogue~\cite{Auchettl:2016qfa} (see also \Ref~\cite{Komossa:2015qya}),  $66$ \td\ candidates  have been identified with various degrees of confidence. Roughly, observations confirm the general theory of tidal disruption, whereas they leave many open questions on the energetics and dynamics of these phenomena -- and possible selection biases in their detection, see \eg\ \Ref~\cite{Kochanek:2016zzg}. 

Interestingly, a subset of all the observed \tds\ show evidence for a relativistic jet, and exhibit a significantly higher luminosity in X-rays.  The best observed jetted \td\ is \sft\ \cite{Burrows:2011dn}; others are \sftt\  \cite{Cenko:2011ys} and \sfttt\  \cite{Brown:2015amy} (the latter being somewhat atypical, see \cite{Auchettl:2016qfa}).  Other transient events   have been proposed to be jetted \tds\ \cite{vanVelzen:2015rlh,Lei:2015fbf}, although their interpretation is less robust.  
It has also been suggested that jetted \tds\ might have have been observed in the past in gamma rays, as a new class of Ultra-Long Gamma Ray Bursts (ULGRBs) \cite{Levan:2015ama}. 

If they indeed generate jets, \tds\ are candidate sources of cosmic rays. This was first investigated by Farrar and Gruzinov \cite{Farrar:2008ex}, who showed that \tds\ naturally meet all the necessary criteria to accelerate protons to energy $E \sim 10^{20}$ eV, and might be sufficiently abundant to account for the observed ultra-high energy cosmic ray flux. Following works \cite{Farrar:2014yla} discussed this result for \tds\ with parameters compatible with the \sft\ event, suggesting that they could explain the recently observed cosmic ray hotspot; see also \Ref~\cite{Pfeffer:2015idq}. 

Under the hadronic hypothesis, jetted \tds\  are also sources of \ns, via proton-photon interactions. A prediction of the  \n\ flux from a \td\   was first published by Wang et al. \cite{Wang:2011ip} for parameters motivated by \sft. The corresponding number of events  at the \ic\ detector was estimated.  A follow up study \cite{Wang:2015mmh} (see also~\cite{Murase:2013ffa}) shows that \tds\ could be hidden \n\ sources, lacking a photon counterpart due to the jet choking inside an envelope made of the debris of the disrupted star\footnote{We do not consider this possibility in our work.}.

The topic of \tds\ as \n\ sources is especially timely. Indeed, \ic\ has discovered a diffuse flux  of high-energy astrophysical neutrinos of dominantly extragalactic origin~\cite{Aartsen:2013jdh}.  So far, no class of objects which could power most of this flux has been identified. 
Specifically,  the contributions from Active Galactic Nuclei (AGN) blazars~\cite{Aartsen:2016lir} and Gamma-Ray Bursts (GRBs)~\cite{Abbasi:2012zw,Aartsen:2014aqy} have been strongly constrained by stacking the information from many different gamma-ray sources. Furthermore, optically thin sources with neutrino production via proton-proton interactions are constrained by the related gamma-ray production from $\pi^0$ decay and its contribution to the diffuse extragalactic gamma-ray backgrounds.  Comparisons with observations have shown that  starburst galaxies cannot be the dominant source powering the diffuse neutrino flux  \cite{Murase:2013rfa,Bechtol:2015uqb}. This favors a photohadronic origin for the \ns, because mechanisms with proton-photon interactions  can reproduce the spectral shape and flavor composition of the observed neutrinos~\cite{Winter:2013cla}. It is plausible that photohadronic sources might be 
hidden in GeV-TeV gamma-rays, because the parameters that cause a high neutrino production efficiency at the same time produce high opacity to gamma-rays at the highest energies~\cite{Murase:2015xka}.

To summarize, if one single class of sources dominates the observed high-energy neutrino flux, it likely obeys the following criteria: (i) the neutrino production occurs by photohadronic interactions, (ii)  the photon counterpart, if any, is more likely to be found in the KeV-MeV energy bands than in the GeV-TeV bands, and (iii) the sources should be abundant enough in the universe, so that each of them individually is sufficiently weak to evade constraints from \n\ multiplets searches~\cite{Kowalski:2014zda,Ahlers:2014ioa,Murase:2016gly}. 

In this work, we study the diffuse flux of \ns\ from \tds , which may address these three criteria. Specifically, we compute the \n\ production from a single \td, and use the existing information on the \td\ demographics to 
 compute the {\it diffuse} flux of \ns\ from \tds.  Parameters motivated by \sft\ observations will be used, with emphasis on the physical scenarios that could reproduce the observed \n\ signal at \ic.  Signatures of a \td\ \n\ flux that could be relevant for future observations will be discussed as well. 

The paper is structured as follows. In \Sec~\ref{sec:tdephysics} the physics of tidal disruption, and the cosmological rate of \tds\ are discussed. 
 \Sec~\ref{sec:jetphysics} presents the details of the numerical calculation of the \n\ flux from a single \td, with results for several illustrative combinations of parameters.  In \Sec~\ref{sec:diffuse} results are given for the diffuse flux expected at Earth from a cosmological population of \tds.  A discussion on the compatibility with the \ic\ data, and future prospects, is given in \Sec~\ref{sec:disc}.

 %%%%%%%%%%%%%%%%%%%%%%%%%%%%%
\section{Physics of \tds\ } 
\label{sec:tdephysics}

\subsection{Tidal disruption and jet formation}

The basic physics of tidal disruption of star by a \bh\ was first discussed in the 1970s and 1980s, in a number of seminal papers  \cite{Hills75,Rees:1988bf,Lacy82,Phinney89}; see \Refs~\cite{DeColle:2012np,Guillochon:2012uc}  for more recent examples.
Here we summarize the main aspects for a star of solar mass and radius, $m=\msun \simeq 1.99\times 10^{33}~{\rm g}$ and $R=\rsun \simeq 6.96\times 10^{10}~$cm. Let $M$ be the mass of the \bh. 

As it moves closer to the \bh, the star can be deformed, and ultimately destroyed by tidal forces. This happens when the star reaches a distance  close enough to the \bh\ so that the force on a mass element (inside the star) due to the self-gravity of the star is comparable to the force produced on the same element by the \bh.  This distance is the tidal radius
\be
r_t = \left( \frac{2 M}{m}\right)^{1/3} R ~\simeq8.8 \times 10^{12} \, {\rm cm}~\left(\frac{M}{10^6 \msun} \right)^{1/3}\frac{R}{\rsun}\left( \frac{m}{\msun} \right)^{-1/3} ~,
\label{rt}
\ee
and the orbital period at such radius is
\be
\tau_t = 2 \pi \left(\frac{r^3_t}{2 M G} \right)^{1/2} 
\simeq 10^4 \, \mathrm{s}~ \left( \frac{R}{\rsun} \right)^{3/2} \left( \frac{m}{\msun} \right)^{-1/2}~.
\label{taut}
\ee

It is useful to compare these quantities with the 
 \bh\ Schwarzschild radius
\be
R_s = \frac{2 M G}{c^2}  \simeq  3 \times 10^{11} \, {\rm cm} \left(\frac{M}{10^6 \msun} \right)~,
\ee 
 and the corresponding time scale
\be
\tau_s \sim 2 \pi R_s/c \simeq 63 \, \mathrm{s} \, \left(\frac{M}{10^6 \msun}\right).
\label{equ:schwtime}
\ee
Here $\tau_s$  is a good approximation of the orbital period at the innermost stable circular orbit, for a Schwarzschild black hole (in the observer's frame, see \eg\ \Ref~\cite{Romero:2014uma}). 

Comparing $r_t$ with $R_s$ shows that the star will be swallowed whole, with no prior disruption, if $M\gta M_{\mathrm{max}} \simeq 10^{8}~{\msun}$.  Here  a more conservative value, $\mmax\simeq 10^{7.2}~\msun$ will be used, motivated by \Ref~\cite{Kochanek:2016zzg}. As will be clear from \Sec~\ref{sub:fluxearth}, our results for the diffuse \n\ flux depend weakly on $\mmax$. 

In the case where disruption occurs, the main phenomenology can be described analytically in terms of basic physics arguments~\cite{Rees:1988bf}. 
About $\sim 1/2$ of the mass of the disrupted star becomes bound to the \bh\ and is ultimately accreted on it.  Therefore, an upper limit to the energy emitted in this event is
\begin{equation}
E_{\mathrm{max}} \sim \msun c^2/2 \simeq 9 \times 10^{53}~{\rm erg} \label{equ:emax} \, ,
\end{equation}
 assuming that the change in the \bh's own internal energy is negligible.  After a dark interval of  ${\mathcal O}(10)$ days -- the time-scale of infall of the tightest bound debris  --  rapid accretion of matter on the \bh\ begins. 
In circumstances where the mass infall rate is sufficiently high --  depending on the detailed dynamics of the stellar debris (see e.g. \Refs~\cite{Shiokawa:2015iia,Dai:2015eua})  -- 
 a flare is generated, with super-Eddington luminosity that declines with time as $ \sim t^{-5/3}$. The flare vanishes rapidly  after a time $\Delta T \sim {\mathcal O}(0.1 - 1)$ yr , when the infall rate drops below the Eddington rate.

Extreme, highly super-Eddington flares are expected if a relativistic jet is launched. The best known jetted \td\ is \sft . Its X-ray flare had an isotropic equivalent luminosity $L_X \simeq 10^{47.5}~{\rm erg~s^{-1}}$   over a time interval $\Delta T \simeq 10^6$ s, for a total energy in X-rays $E_X=  L_X \Delta T \simeq 3 \times 10^{53}$ erg.  Note that the applicable luminosity is arguable, as the average (versus peak) luminosity depends on the time interval considered because the luminosity drops with time, see \eg\ \Ref~\cite{Auchettl:2016qfa}. Here we choose the time window and luminosity which we find most appropriate for neutrino production.

 A minimum variability time $t_v \sim 10^2$ s was observed in the X-ray luminosity, and a Lorenz factor $\Gamma \sim 10$  for the jet was inferred from the data \cite{Burrows:2011dn}. 
The energy $E_X$ is therefore well below the (beaming factor-corrected) maximal isotropic equivalent energy $\sim 2 \Gamma^2 \, E_{\mathrm{max}}$. Parameters motivated by \sft\ are considered typical, as they are overall similar to those of the other well established jetted \td , \sftt\  \cite{Cenko:2011ys}.\footnote{Note however that for \sftt, noise limited the sensitivity to time variability to scales larger than $\sim 10^3$ s.  The smallest time scale of variability observed was at the level of $\sim 10^4$ s \cite{Cenko:2011ys}.}   They were used in \cite{Wang:2011ip} for \n\ flux estimation, and will be used here as well as benchmark (see \Sec~\ref{sub:flux}).

\subsection{Rate of \td}

The cosmological rate of \tds\ is given by the product of the the rate of \tds\ per black hole $\dot N_{\mathrm{TD}}$, the  \bh\ mass function $\phi(z,M)$, defined as the number of black holes per comoving volume and per unit mass at redshift $z$, and the occupation fraction, $f_{\mathrm{occ}}(M)$, which represents the probability that a \bh\ is located at the center of a host galaxy: 
\be
\dot{\rho}(z,M) =  \dot N_{\mathrm{TD}}(M) f_{\mathrm{occ}}(M)\phi(z,M)~.
\label{dotrho}
\ee
 We describe these quantities following mainly Shankar et al. \cite{Shankar:2007zg}, Stone and Metzger \cite{Stone:2014wxa}, and Kochanek \cite{Kochanek:2016zzg}.
In \cite{Shankar:2007zg}  the black hole mass function is calculated for $M \geq 10^5 \msun$, using information from quasar luminosity functions, and estimates of merger rates to model the growth of black holes. Constraints from local estimates of the black hole mass function are taken into account as well.   It is found that  $\phi(z, M)$ declines with $z$ -- roughly as $(1+z)^{-3}$ --  and scales approximately like $M^{-3/2}$ for all $z$ and for $10^5 \msun \lta M \lta 10^{7.5}$. 

The occupation fraction $f_{\mathrm{occ}}$ can be modeled in first approximation as a step function, with $f_{\mathrm{occ}}\simeq 1$ ($f_{\mathrm{occ}}\simeq 0$) above (below) a cutoff mass $\mmin$. Below this mass a number of effects suppress the probability that low mass \bh\ are found in the center of galaxies. For example, a low mass \bh\ is more likely to be ejected from the host galaxy, see e.g.,\cite{Fialkov:2016iib}. 
In \cite{Stone:2014wxa} several possibilities are discussed for the cutoff, with $\mmin \sim (2\times 10^5 - 7 \times 10^6) \msun$.  Instead, in \cite{Kochanek:2016zzg} the entire mass range $10^5 \msun \lta M \lta 10^{7.5} \msun$ is used for the calculation of \td\ rates.  Even lower mass values, down to $M \simeq 10^{4.5}\msun$ are considered in \cite{Fialkov:2016iib}, motivated by  the values of $M$ reported in recent observation of dwarf galaxies \cite{Baldassare15,Baldassare:2016cox}.
We consider the Shankar et al. mass function, extrapolated at $M<10^5\msun$, and use, for illustration, several values in the interval $\mmin=(10^{4.5} - 10^{6.5} )\msun$ for the cutoff mass. 

The rate of tidal disruptions (jetted and non-jetted) per \bh\ decreases weakly with increasing $M$; here we use   $\dot N_{\mathrm{TD}} \simeq 10^{-3.7} (M/10^6 \msun)^{-0.1}~{\rm yr^{-1}}$  \cite{Kochanek:2016zzg}, which is  close to the upper limit obtained from the ASAS-SN data \cite{Holoien:2015pza}.  We consider only the {\it total} rate of disruptions per \bh, neglecting their distribution in the mass of the disrupted star $m$.  As shown in \cite{Kochanek:2016zzg}, this distribution ranges in the interval $m \sim (0.1 - 2)\msun$, with $m \simeq 0.3 \msun$ a typical value. Variations of $m$ in this interval would only produce weak effects (less than a factor of $\sim2$) in our calculations (see \eqs~(\ref{rt}) and (\ref{taut})). These effects are subdominant  compared to those of $M$ varying over two orders of magnitude, and therefore they are neglected here.\footnote{Note also that the mass of the disrupted star, $m$, does not directly enter our calculation, where we use observed values of the X-ray luminosity and total energy, $L_X$ and $E_X$, as inputs; see \Sec~\ref{sec:comp}. }

\begin{figure}[tbp]
\centering
\includegraphics[width=0.48\textwidth]{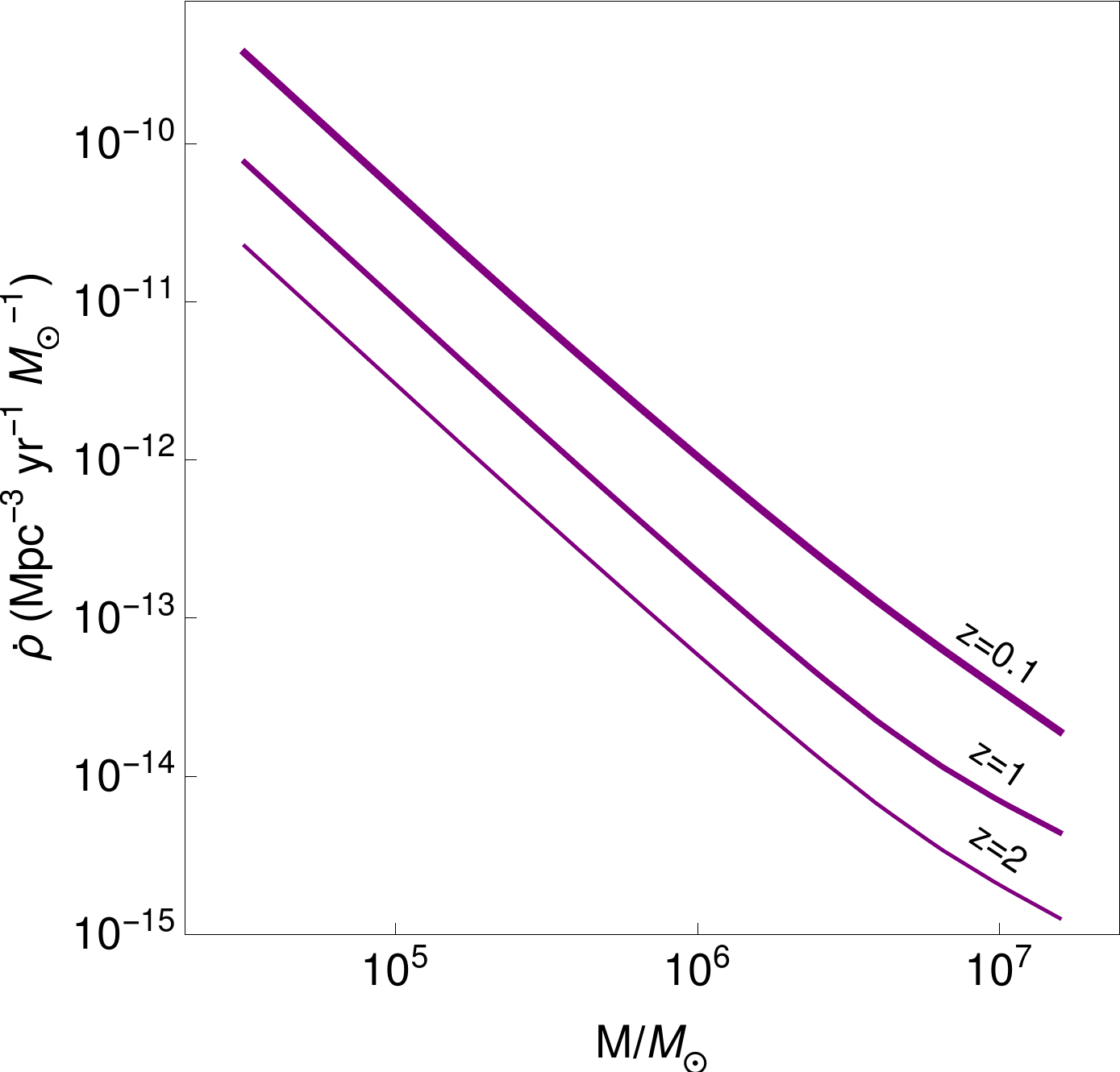} \hspace*{0.02\textwidth}
\includegraphics[width=0.48\textwidth]{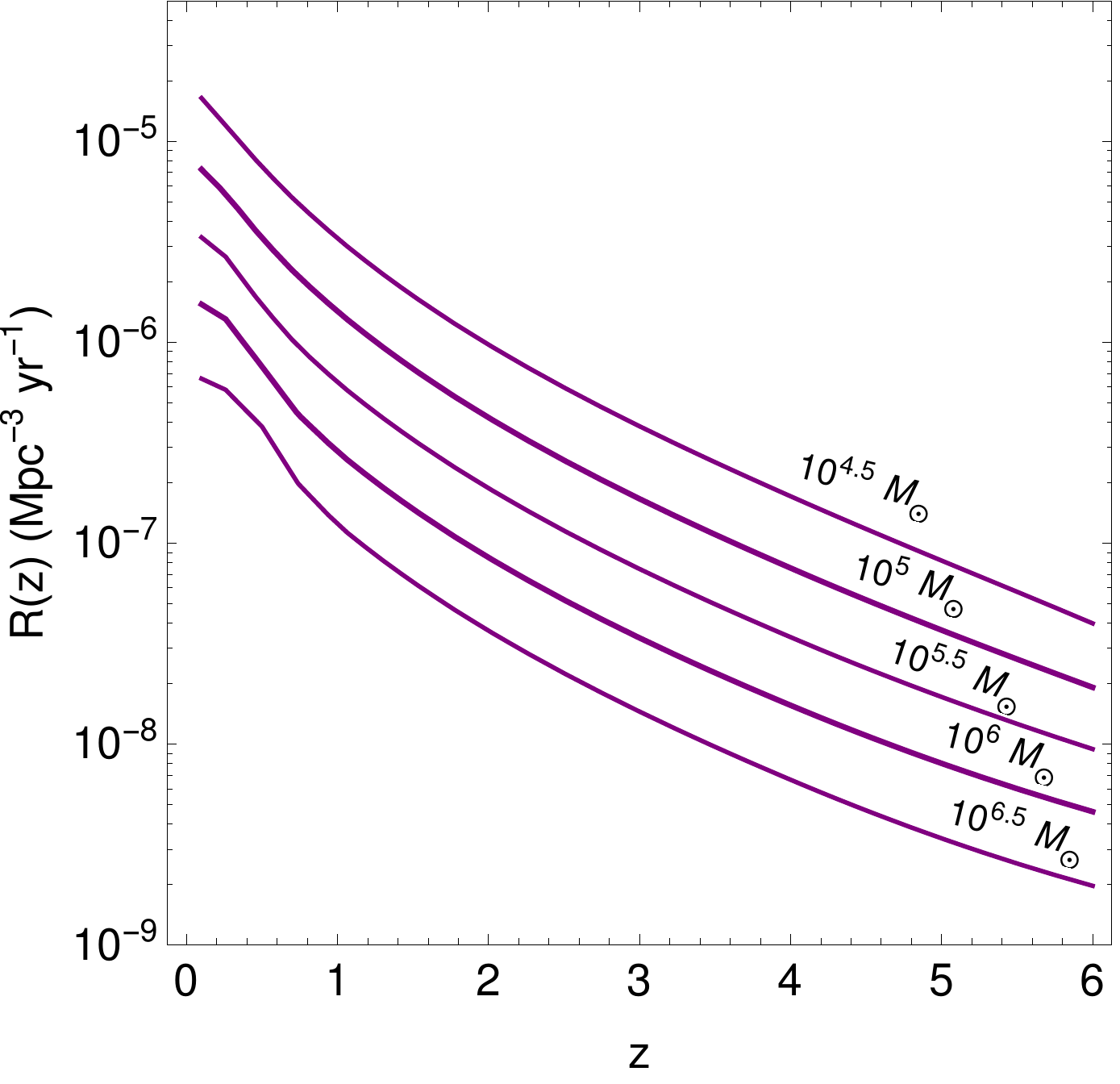}
\caption{ {\it Left panel: }  The differential rate of \tds\ (jetted and not jetted)  $\dot \rho(z,M)$ as a function of $M$ for selected values of $z$ (labels on curves).  The mass interval $[\mmin,\mmax]=[10^{4.5} \msun, 10^{7.2} \msun]$ is used here.
 {\it Right panel: } The total volumetric rate of \tds, $R(z)$, for $\mmax=10^{7.2} \msun$ and different values of $\mmin$ (labels on curves). 
 }
\label{fig:tderates}
\end{figure}

\figu{tderates} shows the differential \td\ rate, $\dot{\rho}(z,M)\propto M^{-1.6}$ (left), and the total rate $R(z)=\int^{\mmax}_{\mmin} \dot{\rho}(z,M) dM$, as a function of $z$ (right).  As expected, $R(z)$ is dominated by the lowest mass \bh, decreasing by a factor $\sim 10^{0.6} \simeq 4$ when $\mmin$ is increased by an order of magnitude.

Consider the effective rate of observable jetted \tds\ $\tilde R$, which can be estimated as $\tilde R = R \, \eta/(2 \Gamma^2)$ with the beaming factor $1/(2 \Gamma^2)$ and the fraction $\eta$ of all \td s producing a jet. Using $\Gamma \simeq 10$~\cite{Burrows:2011dn} and $\eta \sim 0.1$~\cite{Burrows:2011dn}, the suppression factor between observable jetted and all \td s can be estimated to be $\sim 5 \cdot 10^{-4}$. Consequently, the local rate of observable jetted \td s is expected to be $\tilde R(0) \simeq 0.35 - 10 \, \mathrm{Gpc^{-3}} \, \mathrm{yr^{-1}}$, depending on $\mmin$. Note that this rate is still subject to possible selection biases if one compares it to data.

It is interesting to compare the expected jetted \td\ rate $\tilde R(0)$ to constraints from current \ic\ data 
\cite{Kowalski:2014zda,Ahlers:2014ioa,Murase:2016gly}. For example, \Ref~\cite{Ahlers:2014ioa}  discusses the case of transient sources, under the assumption that they contribute to most of the astrophysical \n\  flux observed at \ic.  The main result is that rare but powerful transients, with a local rate  $\tilde R(0) < 10 \, \mathrm{Gpc}^{-3} \, \mathrm{yr}^{-1}$,  can be  excluded within five years of operation (corresponding to present data) from the non-observation of multiplets. These bounds apply to short transients (like Gamma Ray Bursts); they relax somewhat for longer lived sources like \tds .\footnote{The other references come to similar conclusions, with some dependence on source evolution history, spectral shape \etc }
However, it is already evident from that estimate that a diffuse neutrino flux from \tds\ describing IceCube data must be dominated by the low mass part of the \bh\ mass function in order to avoid the tension with these constraints.  
 A next generation instrument~\cite{Aartsen:2014njl} will be more sensitive, being able to identify sources that are more frequent but less bright. A bound might be as strong as $\tilde R(0) < 10^3 \, \mathrm{Gpc}^{-3} \, \mathrm{yr}^{-1}$, which can clearly test the \td\ hypothesis. We will discuss another way to test of the TDE hypothesis, using the flux flavor composition, in \Sec~\ref{sub:detection}.  

%

%%%%%%%%%%%%%%%%%%%%%%%%%%%%%%%%%%%%%%%%%%
\section{Neutrino production in a \td-generated jet} 
\label{sec:jetphysics}

\begin{table}[t]
\caption{Parameters used in this work,  unless noted explicitly otherwise. These parameters apply to the \bh\ frame.
\label{default}
} 
\begin{center}
\begin{tabular}{|c|lr|}
\hline
Symbol & Definition & Standard value \\
\hline
$t_v$ &  Variability timescale & $10^2 \, \mathrm{s}$  \\
$\Gamma$ &  Lorentz factor  &  10 \\
$\xi_p$ &  Baryonic loading (energy in protons versus X-rays)  & 10  \\
$\xi_B$ & Magnetic loading (energy in magnetic field versus X-rays) &  1 \\
$k_p$ &  Proton spectral index  & $2 $ \\
$E_X$ &  Isotropic equivalent energy in X-rays  & $3 \cdot  10^{53} \, \mathrm{erg} $ \\
$\Delta T$ & Duration of X-ray flare  & $10^6$ s  \\
$\varepsilon_{X,\mathrm{br}}$ & Observed X-ray break energy   & 1 keV \\
$\alpha$ & Lower X-ray spectral index $\varepsilon < \varepsilon_{X,\mathrm{br}}$    & $2/3$ \\
$\beta$ & Higher X-ray spectral index $\varepsilon > \varepsilon_{X,\mathrm{br}}$    & $2$ \\
$\eta$ & Fraction of \tds\ with jet formation (used for diffuse flux) & 0.1 \\
\hline
\end{tabular}
\label{tab:parameters}
\end{center}
\end{table}

\subsection{Photohadronic processes and neutrino emission}
\label{sec:comp}

For the computation of the neutrino flux from a single TDE, we follow the relativistic wind description in \Ref~\cite{Wang:2011ip}. We apply however methods as they have been used in state-of-the-art calculations for relativistic winds in Gamma-Ray Bursts before~\cite{Baerwald:2010fk,Baerwald:2011ee,Hummer:2011ms} using the NeuCosmA software. A comparison between the numerical computation used in this study and the analytical estimate can be found in \App~\ref{app:analytical}. Our standard parameter values are summarized in \Tab~\ref{default}. We note that the  approach in this section could be also easily applied to \td\ stacking analyses, for which the required input is listed in \Tab~\ref{default} (except from $\eta$, but including $z$); in fact, a very similar method has been used for Gamma-Ray Burst stacking in \Ref~\cite{Adrian-Martinez:2013dsk}.

 The photon spectrum is assumed to fit the observed spectral energy distributions of TDEs described as a broken power law with a spectral break,  parameterized in the shock rest frame (SRF) by (we use primed quantities for the SRF)
 \begin{equation}
 	N'_{\gamma}(\varepsilon') = C'_{\gamma} \cdot \left\{ \begin{array}{ll} \left( \frac{\varepsilon'}{\varepsilon'_{X,\mathrm{br}}} \right)^{-\alpha} & \varepsilon'_{X,\text{min}} \leq \varepsilon' < \varepsilon'_{X,\mathrm{br}} \\ \left( \frac{\varepsilon'}{\varepsilon'_{X,\mathrm{br}}} \right)^{-\beta} & \varepsilon'_{X,\mathrm{br}} \leq \varepsilon' < \varepsilon'_{X,\text{max}} \\ 0 & \text{else} \end{array} \right. \quad , \label{equ:targetphoton}
 \end{equation}
 where $C'_{\gamma}$ is a normalization factor. Typical values can be found in \Tab~\ref{default}, where $\varepsilon'_{X,\mathrm{br}}=\varepsilon_{X,\mathrm{br}} (1+z)/\Gamma$, and $\varepsilon'_{X,\text{min}}$ and $\varepsilon'_{X,\text{max}}$ can be translated from the observed energy band correspondingly. We use the Swift energy band with $\varepsilon_{X,\text{min}} \simeq 0.4 \, \mathrm{keV}$ and $\varepsilon_{X,\text{max}} \simeq 13.5 \, \mathrm{keV}$~\cite{Burrows:2011dn} to define the target photon spectrum, unless noted otherwise. Note that one may define a bolometric correction to that (such as one may extend the target photon spectrum beyond that range), but the increase of the neutrino flux would be small as long as the break energy was sufficiently well covered.

 The proton spectrum is assumed to be a cut-off power law with a spectral index $k_p \simeq 2$ expected from Fermi shock acceleration
 \begin{equation}
 	N'_p(E_p') = C'_p \cdot \left\{ \begin{array}{ll} \left( \frac{E_p'}{\mathrm{GeV}} \right)^{-k_p} \cdot \exp \left( - \frac{E_p'^{2}}{E'^{2}_{p,\text{max}}} \right) &  E_p' \; \ge \;  E'_{p,\text{min}}  \\ 0 & \text{else} \end{array} \right. . \label{equ:targetproton}
 \end{equation}
The maximal proton energy $E'_{p,\text{max}}$ is determined automatically by balancing the acceleration rate with synchrotron loss and adiabatic\footnote{The adiabatic cooling timescale is chosen to be similar to the dynamical timescale, which means that it is implied that the dynamical timescale can limit the maximal energy.} cooling rates  and comes from the cutoff from acceleration, and we choose $E'_{p,\text{min}} \simeq 1 \, \mathrm{GeV}$. However, for the neutrino production in TDEs the maximal proton energy is not so important as long as $E'_{p,\text{max}} \gtrsim 10^8 \, \mathrm{GeV}$, because the magnetic field effects on the pions and muons will dominate the maximal neutrino energies and the energy budget only logarithmically depends on $E'_{p,\text{min}}$ and $E'_{p,\text{max }}$ for $k_p=2$.  Consequently, the chosen shape of the (super-exponential) cutoff, which may be relevant for the description of ultra-high energy cosmic rays, does not have any impact on the neutrino flux computation. 

The  isotropic  equivalent energy $E_X$ (in $\text{erg}$) is given in the SMBH frame\footnote{A clarification is due on the definition of frames of reference used here. For brevity, the wording ``\bh\ frame'' will be used to indicate a frame of reference of an observer  at rest with respect to the \bh\ and  located at a distance $L$ from it such that $R_s \ll L \ll c/H_0$. Instead, ``observer's frame'' indicates the frame of reference of Earth. Energies in the two frames differ by redshift effects.  } as 
\begin{equation}
	E_{X} = \frac{4\pi \, d_L^2}{(1+\textit{z})} \; S_{X}  \quad .
\label{equ:eisobol}
\end{equation} 
in terms of the X-ray fluence $S_X$ (in units of $\mathrm{erg \, cm^{-2}}$). Note again that this fluence is assumed to be measured in the energy band from 0.4 to 13.5~keV . The isotropic energy $E_X$ can be obtained from the X-ray luminosity by $E_X=L_X \Delta T_{\mathrm{obs}}/(1+z)$. Here we see already one known subtlety: redshift enters here because the observed duration $\Delta T_{\mathrm{obs}}$ is defined in the observer's frame and $E_X$ and $L_X$ in the \bh\ frame. For the computation of the diffuse flux, it will be most convenient to define all quantities in the \bh\ frame, including $\Delta T$, $t_v$, $\varepsilon_{X,\text{br}}$, $\varepsilon_{X,\text{min}}$, and $\varepsilon_{X,\text{max}}$, which means that all TDEs with the same parameters will be alike in the \bh\ frame.  
Practically, we implement that by computing the neutrino fluxes for a \td\ that takes place at a very small  $z\ll1$  (where the oberver's frame is basically identical with the \bh\ frame). 
This computation gives the neutrino fluence, \ie, the number of \ns\ of a given flavor that reach Earth per unit energy per unit area, $ F_\alpha(E)$ ($\alpha=e,\mu,\tau$). For future use, it is convenient to also consider the number of produced neutrinos  of a given flavor (after oscillations)  per unit energy $Q_\alpha(E)$, which is related to the fluence by $F_\alpha(E)=Q_\alpha (1+z)^3/(4\pi \, d_L^2)$ (which is $F_\alpha(E)=Q_\alpha /(4\pi \,L^2)$ for small $z$ with the lookback distance $L \simeq d_L$). 
From $F_\alpha$ and $Q_\alpha$, the fluence of \ns\ for a generic \td\ at any redshift $z$ is obtained by the appropriate re-scaling.  
We checked that the difference between the two methods (all parameters alike in \bh\ frame versus oberver's frame) is small.

The isotropic energy  can be easily boosted into the SRF by $E'_X= E_X/\Gamma$.
Assuming that the emitted photons are coming from synchrotron emission of electrons (or mainly interact with electrons), the amount of energy in electrons and photons should be roughly equivalent. 
In a baryonically dominated relativistic wind, we have
\begin{equation}
	 E'_p  \simeq E'_X  \, \xi_p \, \,
\label{equ:eiso}
\end{equation}
 where $\xi_p$ is the ratio between proton and X-ray energy -- referred to as ``baryonic loading''. 

We compute the photon and proton densities in the SRF defining  an ``isotropic volume'' $V'_{\mathrm{iso}}$, which is the volume of the interaction region in the source frame assuming isotropic emission of the engine.
 Thus, the assumption of isotropic emission will cancel in the density. Similarly, $V'_{\mathrm{iso}}$ is an equivalent volume in the SRF where only the radial direction is boosted, which is given by
\begin{equation}
	V'_{\mathrm{iso}} = 4\pi \, R_C^2 \cdot \Delta d' 
	\label{equ:visoISM}
\end{equation}
with shell width $\Delta d' \simeq \Gamma c t_v/(1+z)$ obtained from the variability timescale, and the collision radius $R_C \simeq  2 \, \Gamma^2 \, c \, t_v/(1+\textit{z})$.
From \equ{visoISM} we can then estimate the size of the interaction region as $V'_{\mathrm{iso}}  \propto \Gamma^5 t_v^3$, which means that it strongly depends on the $\Gamma$ factor.

Because of the intermittent nature of TDEs, the total fluence is assumed to be coming from $N \simeq \Delta T/t_v$ such interaction regions.
Now one can determine the normalization of the photon spectrum in \equ{targetphoton} from
\begin{equation}
 \int  \, \varepsilon' \, N'_{\gamma}(\varepsilon') \mathrm{d}\varepsilon' = \frac{E'_X}{N \, V'_{\text{iso}}} \quad . \label{equ:photonorm}
\end{equation}
if one assumes that the target photons can escape from the source. Note that $E_X/N \simeq L_X t_v$, which means that one can use $L_X$ equivalently to define the target photon density or pion production efficiency -- as we do in the analytical approach in \App~\ref{app:analytical}.

Similarly, one can compute the normalization of the proton spectrum in \equ{targetproton} by
\begin{equation}
\int \, E'_p \, N'_p(E'_p) \, \mathrm{d}E'_p =  \xi_p \cdot \frac{E'_X}{N \, V'_{\text{iso}}}  \, .
\label{equ:protonorm}
\end{equation}
Given that the ratio between magnetic field and X-ray energies is $\xi_B$, one has in addition\footnote{With this definition of $\xi_B$, the magnetic loading is slightly different from \Ref~\cite{Wang:2011ip}, who define the magnetic energy with respect to the wind luminosity (which is a factor of three higher than the radiated energy). Their magnetic loading is therefore effectively a factor of three higher than ours.}
\begin{equation}
 U'_B = \xi_B  \cdot \frac{E'_X}{N \, V'_{\text{iso}}} \quad \text{or} \quad B' = \sqrt{8\pi  \cdot \xi_B \cdot \frac{E'_X}{N \, V'_{\text{iso}}}}  .
\label{equ:B}
\end{equation}

Once the proton and photon densities and the magnetic field are determined, the rest of the computation is straightforward. We solve the time-dependent differential equation system for the pion and consequent muon densities, including photo-meson production based on SOPHIA~\cite{Mucke:1999yb} (with an updated  method similar to \Ref~\cite{Hummer:2010vx} and first used in \Ref~\cite{Boncioli:2016lkt}). We also include the  helicity-dependent muon decays~\cite{Lipari:2007su} and the leading kaon production mode. The radiation processes of the secondary pions, muons and kaons include synchrotron losses, adiabatic losses, and escape through decay, which lead to characteristic cooling breaks different for pions, muons, and kaons, and a transition in the flavor composition~\cite{Kashti:2005qa}; see \App~\ref{app:analytical} for an analytical discussion.

\subsection{Neutrino fluence from a tidal disruption event}
\label{sub:flux}

\subsubsection{Modeling the jet: inputs and assumptions}
\label{subsub:scalings}

Considering that the masses of the black holes responsible for \tds\ may vary over more than two orders of magnitude, it is natural to expect a certain degree of diversity in the jetted \tds .  Here 
 we estimate  how certain parameters of the jet may depend on the \bh\ mass.  For the sake of generality, we choose parameter scalings that either have an observational basis, or a direct connection to fundamental physics.  
 
 Let us first discuss three parameters that most influence the \n\ flux, the minimum variability time $t_v$, the Lorenz factor $\Gamma$, and the luminosity $L_X$.    Observations of Active Galactic Nuclei indicate a mild dependence of $\Gamma$ on $M$ which is best fit by \cite{Chai:2012ns} 
\be 
\Gamma = \left(\frac{M}{10 \msun}\right)^{0.2} ~.
\label{equ:gammafun}
\ee
This corresponds to $\Gamma\sim 6,10,$ and $16$ for $M = 10^5, 10^6,$ and $10^7 \msun$, respectively, which are compatible with observational estimates for \sft.  The relationship in \equ{gammafun}  is consistent \cite{Chai:2012ns} with the magnetically arrested accretion flow model \cite{Tchekhovskoy11,McKinney12}, which predicts $\Gamma$ to depend on the square of the \bh\ spin, which in turn increases with $M$.   Still, it is not known if \equ{gammafun} applies to the broader set of galaxies (most of them not hosting an active nucleus) of interest here. Therefore, it should be considered as a mere possibility, although theoretically substantiated. 

For the minimum variability time, $t_v$, it is reasonable to make the hypothesis that it be related to the smallest possible time scale available in a black hole, the Schwarzschild ``time" $\tau_s$, \equ{schwtime}.  Typical values of $\tau_s$ are consistent with the variability seen in \sft, and indeed the hypothesis of a connection to the Schwarzschild time is used in interpretations of \sft\ data to infer the mass of the parent \bh\ \cite{Burrows:2011dn}. 

Lastly, we can expect some dependence of $L_X$ on $M$.  Combined data on jetted and non-jetted \tds\ are well fit by a luminosity function that scales like the inverse square of $L_X$ \cite{Sun:2015bda}:\footnote{One can  substitute the integral over $M$ by one in $L_X$ in the diffuse flux \equ{jtde} for $L_X \propto M^\alpha$, in which case one observes that ($\dot \rho(M)/\Gamma^2 (M)$ corresponds to the luminosity distribution function.}
\be
\frac{\dot \rho(M)}{\Gamma^2(M)} \propto L^{-2}_X~,
\label{equ:lumifun}
\ee
which implicitly gives a scaling $L_X= L_X(M) \propto \Gamma \dot \rho^{-1/2}$.  
Eq. (\ref{equ:lumifun}) is just a a possibility. Indeed, it is also possible that X-ray flares from jetted TDEs do not follow the same trend as the ones from non-jetted TDEs (see Fig.~11 of \Ref~\cite{Sun:2015bda}). 
There are also different scalings, resulting from  theoretical relationships between $X$-ray luminosity and \bh\ mass (\eg\ \Ref~\cite{DeColle:2012np,Guillochon:2012uc}) such as coming from a possible connections between the peak $X$-ray luminosity and the  rate of accretion. 

For the purpose of illustrating possible different degrees of dependence of the \n\ flux on the \bh\ mass, $M$, we present results for four scenarios (see Table~\ref{tab:cases}):

\begin{itemize}
\item {\it Base case}.  Here no dependence  on $M$  is considered at all, and a single set of jet parameters is assumed to describe all jetted \td.  The parameters are the same as in \cite{Wang:2011ip},
 see  \Tab~\ref{default}. 

\item  {\it Weak scaling} case.  Here the weak dependence of $\Gamma$ on $M$, motivated by AGN observations, is included, \equ{gammafun}. All other parameters are as in the base case, except the variability time, which is taken to be   $t_v=10^3$ s (which is more conservative for neutrino production).  This value is meant to illustrate a different  possibility, relative to the base value in Table \ref{default}, and is motivated by the \emph{ median} (rather than minimum) scale of time variability observed in \sft. 

\item {\it Strong scaling} case. Here both $\Gamma$ and $t_v \sim \tau_s$ scale with $M$ as  given in \eqs~(\ref{equ:schwtime}) and (\ref{equ:gammafun}). This means that, in addition to $\Gamma$ scaling in the Weak case, it is assumed that the time variability of the jet is correlated with the period of the lowest stable orbit of the star disrupted by the \bh .

\item {\it Lumi scaling} case.  Here the same scalings as the Strong case are used, and additionally the scaling of $L_X$ is included, as in \equ{lumifun}. Explicitly, considering that $\dot \rho(M) \propto M^{-1.6}$ and $\Gamma^2 \propto M^{0.4}$ (\cf, \equ{gammafun}), \equ{lumifun} implies that   $L_X \propto M$.  We therefore take $L_X = 3 \cdot 10^{47} \,  M/(10^6 \msun) \, \mathrm{erg \, s^{-1}}$, such that the luminosity is the same as in the Strong case for the benchmark \bh\ mass $M=10^6 \msun$.

\end{itemize}

\begin{table}[t]
\caption{Our standard scaling scenarios.}
\label{tab:cases}
\begin{center}
\begin{tabular}{|l|cccc|}
\hline
Case & $t_v$ [s] & $\Gamma$ & $L_X$ [erg/s] & Reference(s) \\
\hline
Base case & $10^2$ & $10$ & $3 \cdot 10^{47}$ & \Tab~\ref{default}, \Ref~\cite{Wang:2011ip} \\
Weak scaling case & $10^3$ & $\Gamma = \left( \frac{M}{10 \msun} \right)^{0.2}$ & $3 \cdot 10^{47}$ & \equ{gammafun}, \Ref~\cite{Chai:2012ns}   \\
Strong scaling case & $t_v \simeq 63  \,  \frac{M}{10^6 \msun} $  & $\Gamma = \left( \frac{M}{10 \msun} \right)^{0.2}$  & $3 \cdot 10^{47}$ & \eqs~(\ref{equ:schwtime}), (\ref{equ:gammafun}), \Ref~\cite{Chai:2012ns}  \\
Lumi scaling case &  $t_v \simeq 63  \, \frac{M}{10^6 \msun}$  &  $\Gamma = \left( \frac{M}{10 \msun} \right)^{0.2}$  & $3 \cdot 10^{47} \,  \frac{M}{10^6 \msun}$ &  \Ref~\cite{Sun:2015bda} (see text) \\
\hline
\end{tabular}
\end{center}
\end{table}

The general effect of the scalings proposed here can be understood by embedding \eqs~(\ref{equ:schwtime}),  (\ref{equ:gammafun}), and (\ref{equ:lumifun}) in the analytical formalism in \App~\ref{app:analytical}.  Considering that $\Gamma$ and/or $t_v$ increase with $M$, we expect: 
(i) a decrease of the pion production efficiency $f_{p\gamma}$, and therefore of the \n\ production, with the increase of $M$. Indeed, $f_{p\gamma} \propto \Gamma^{-4} t^{-1}_v$  (\equ{fpig}), which implies  $f_{p\gamma} \propto M^{-0.8}$ ($f_{p\gamma} \propto M^{-1.8}$) in the Weak (Strong) case; similar results can be found from \equ{visoISM}.
This means that smaller \bh\ masses imply higher pion production efficiencies.
(ii) An increase of the proton, pion and muon break energies with $M$, resulting in a hardening of the \n\ spectrum. This  is because these energies scale as, respectively, $E_{p,\mathrm{br}}\propto \Gamma^2$, $E_{\pi,\mathrm{br}}\propto \Gamma^4 t_v$ , $E_{\mu,\mathrm{br}}\propto \Gamma^4 t_v$ (\eqs~(\ref{epsilonb}), (\ref{zetapi}) and (\ref{emubreak})).   This means that $E_{p,\mathrm{br}}\propto M^{0.4}$ in both scalings, and $E_{\pi,\mathrm{br}},E_{\mu,\mathrm{br}}, \propto M^{0.8}$ ($E_{\pi,\mathrm{br}},E_{\mu,\mathrm{br}}, \propto M^{1.8}$) in the Weak (Strong) case. 

When the dependence of $L_X$ on $M$ is included (Lumi case), the neutrino flux roughly scales as $\phi \propto L_X \, f_{p \gamma} \propto M^{0.2}$, thus increasing  slightly  with $M$, contrary to the other scenarios considered here.

All our \n\ flux calculations include \n\ oscillations in vacuum, with the exception of \figu{numericalvsanalytical} in \App~\ref{app:analytical}.  We have checked that a possible envelope ahead of the jet, caused by debris of the disrupted star, is sufficiently thin that matter-driven flavor conversion is negligible. Due to the extremely long propagation distance, only the effect of averaged oscillations is observable in a detector, and the corresponding flavor conversion probabilities depend only on the mass-flavor mixing matrix, $U$: $P(\nu_\alpha \rightarrow \nu_\beta)=\sum_{i=1}^3 |U_{\alpha i}|^2|U_{\beta i}|^2$, where $\alpha,\beta=e,\mu,\tau$ and $i=1,2,3$ runs over the \n\ mass eigenstates.  We use the standard parameterization of the mixing matrix, with the following values of the mixing angles: $\sin^2 \theta_{12}=0.308$, $\sin^2 \theta_{23}=0.437$, $\sin^2 \theta_{13}=0.0234$ (see \eg\ \Ref~\cite{Olive:2016xmw}). 

\subsubsection{Neutrino fluence at Earth}

\begin{figure}[tbp]
\centering
\includegraphics[width=0.42\textwidth]{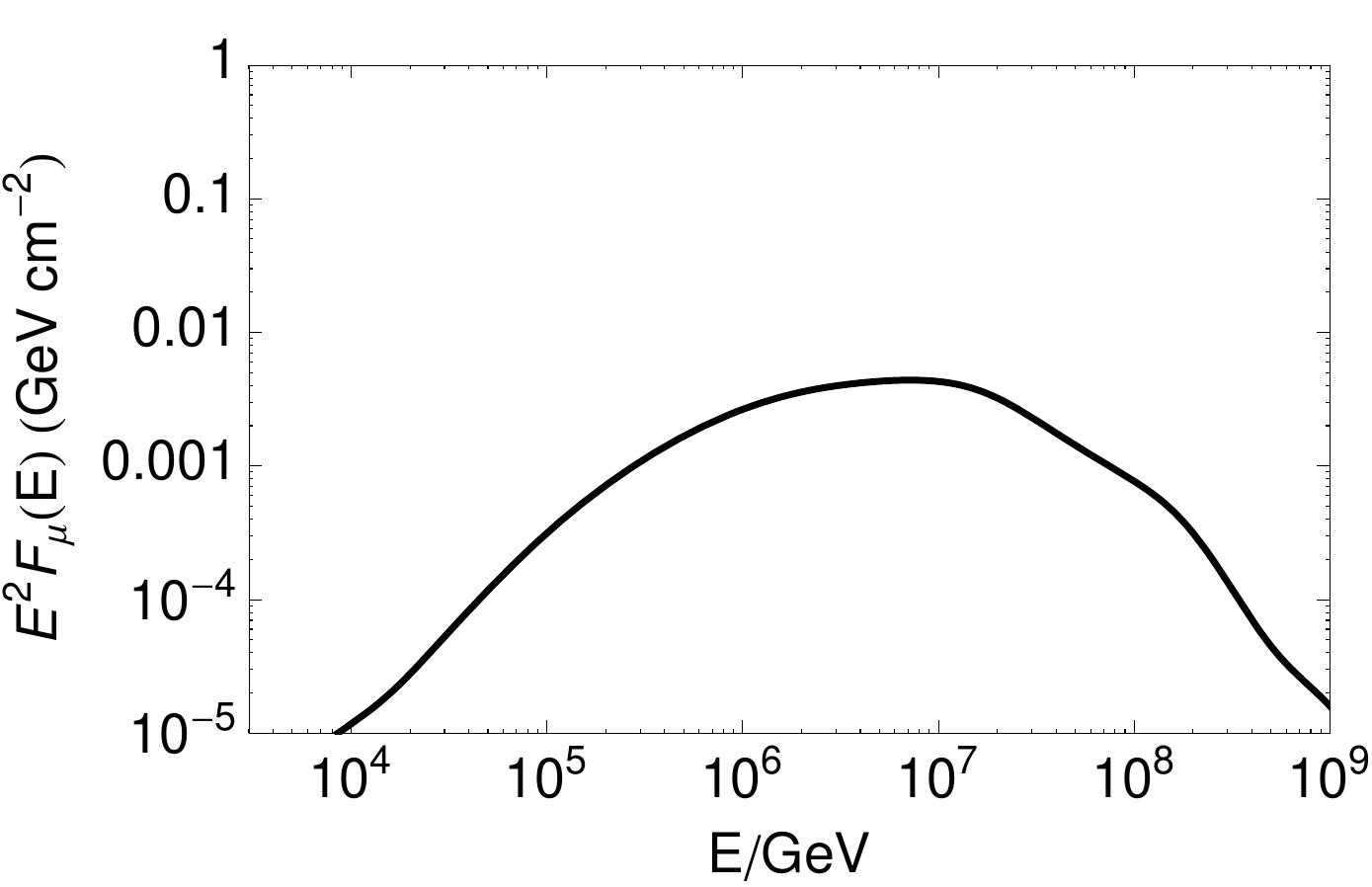}
\includegraphics[width=0.42\textwidth]{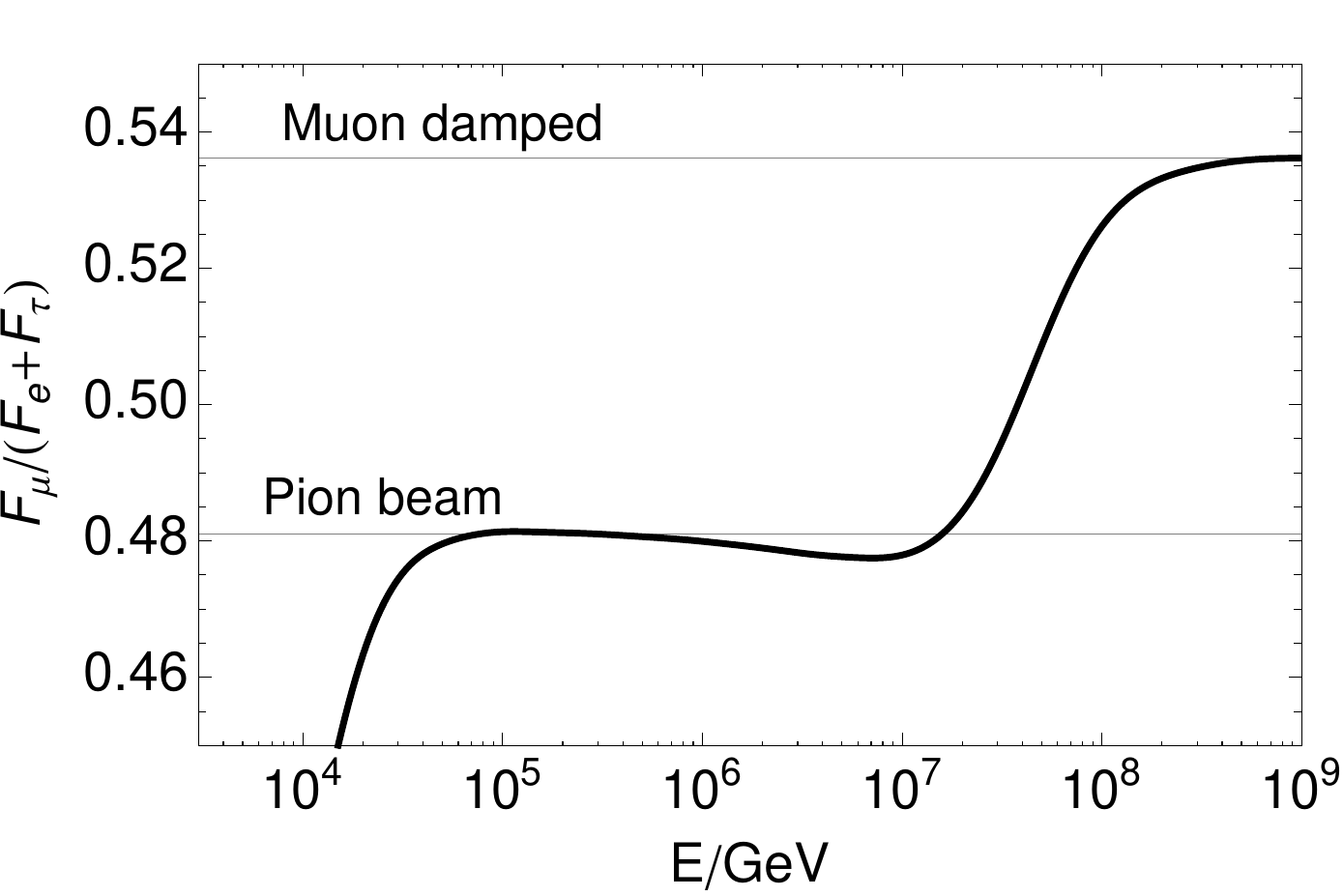}
\includegraphics[width=0.42\textwidth]{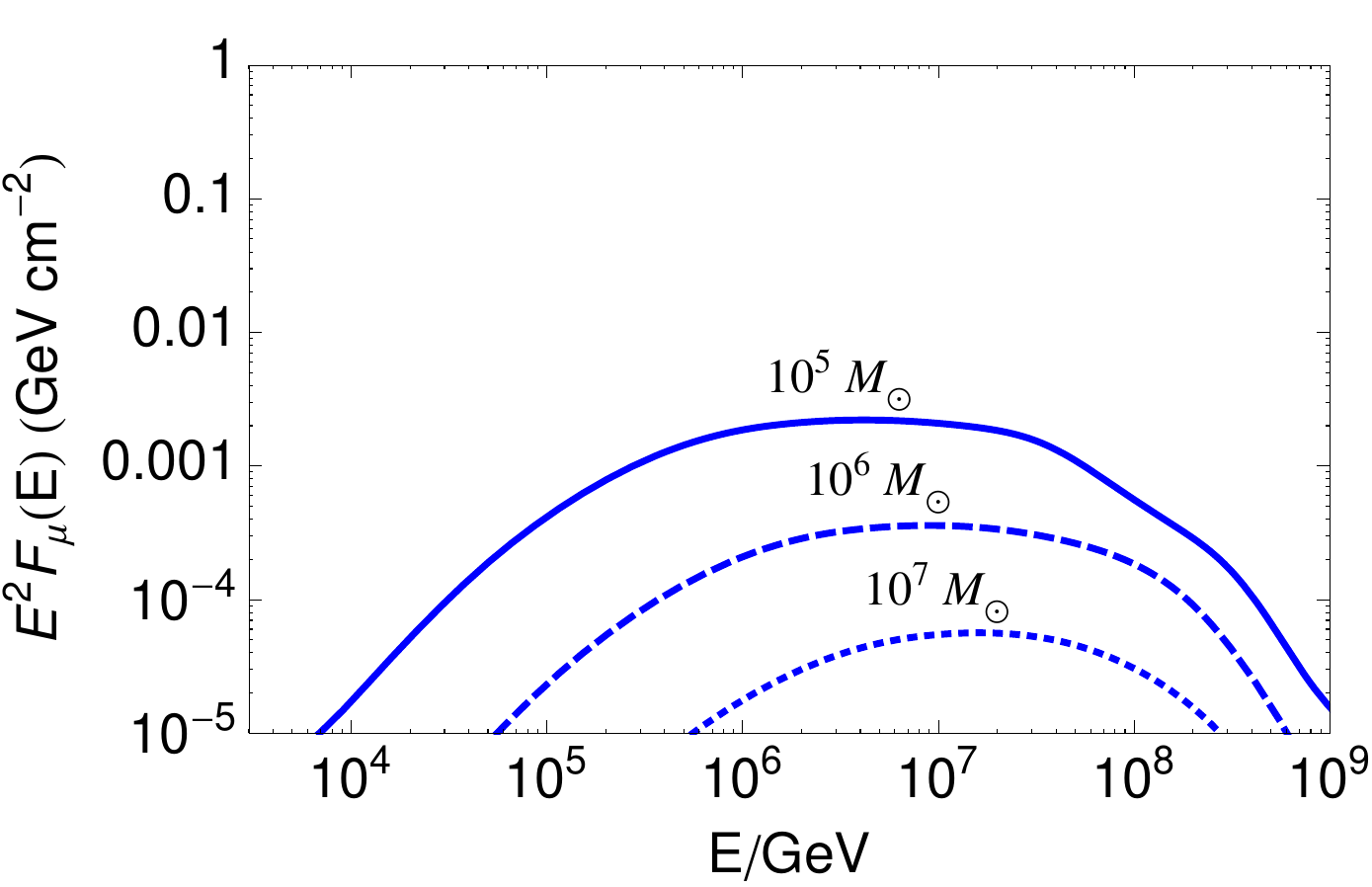}
\includegraphics[width=0.42\textwidth]{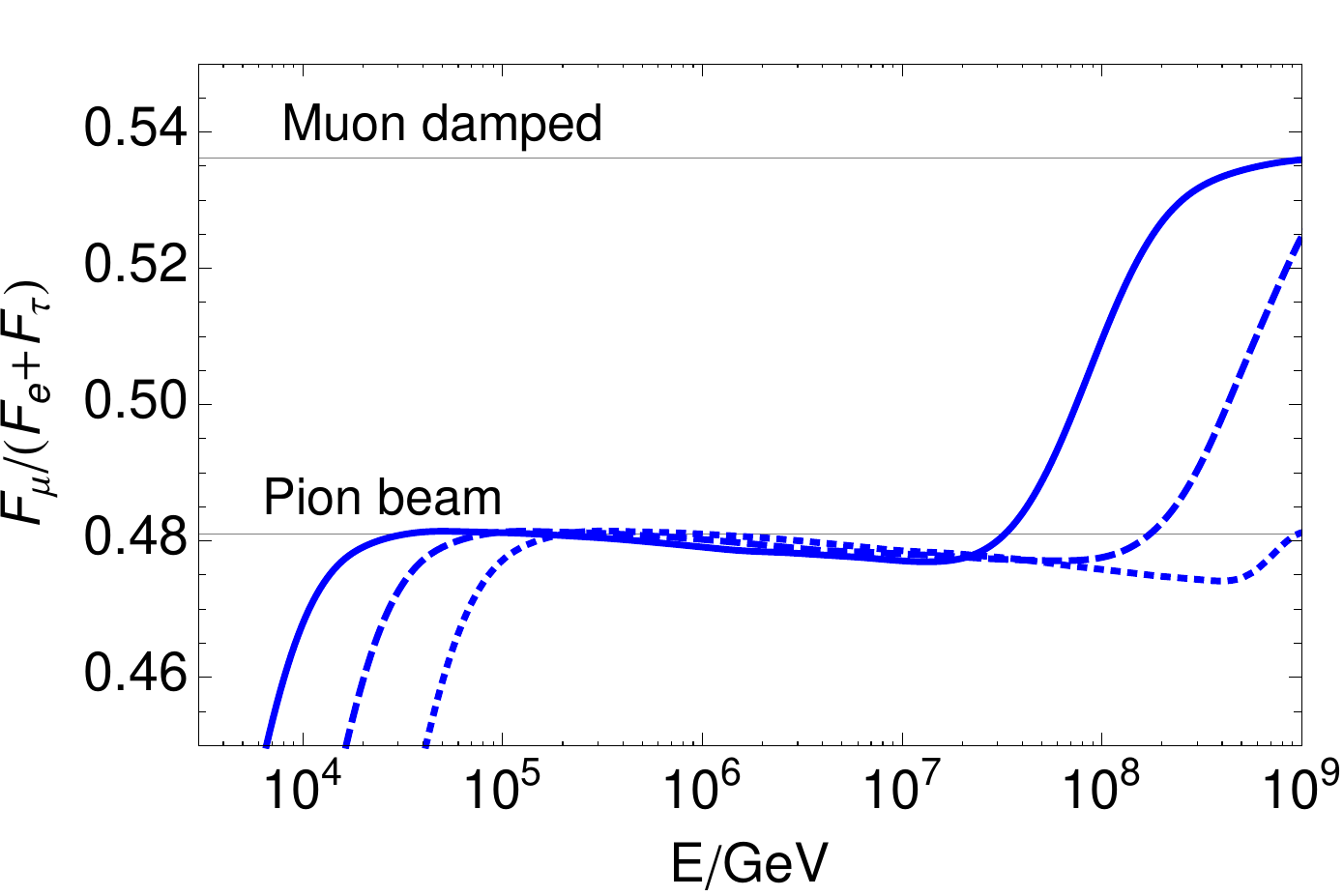}
\includegraphics[width=0.42\textwidth]{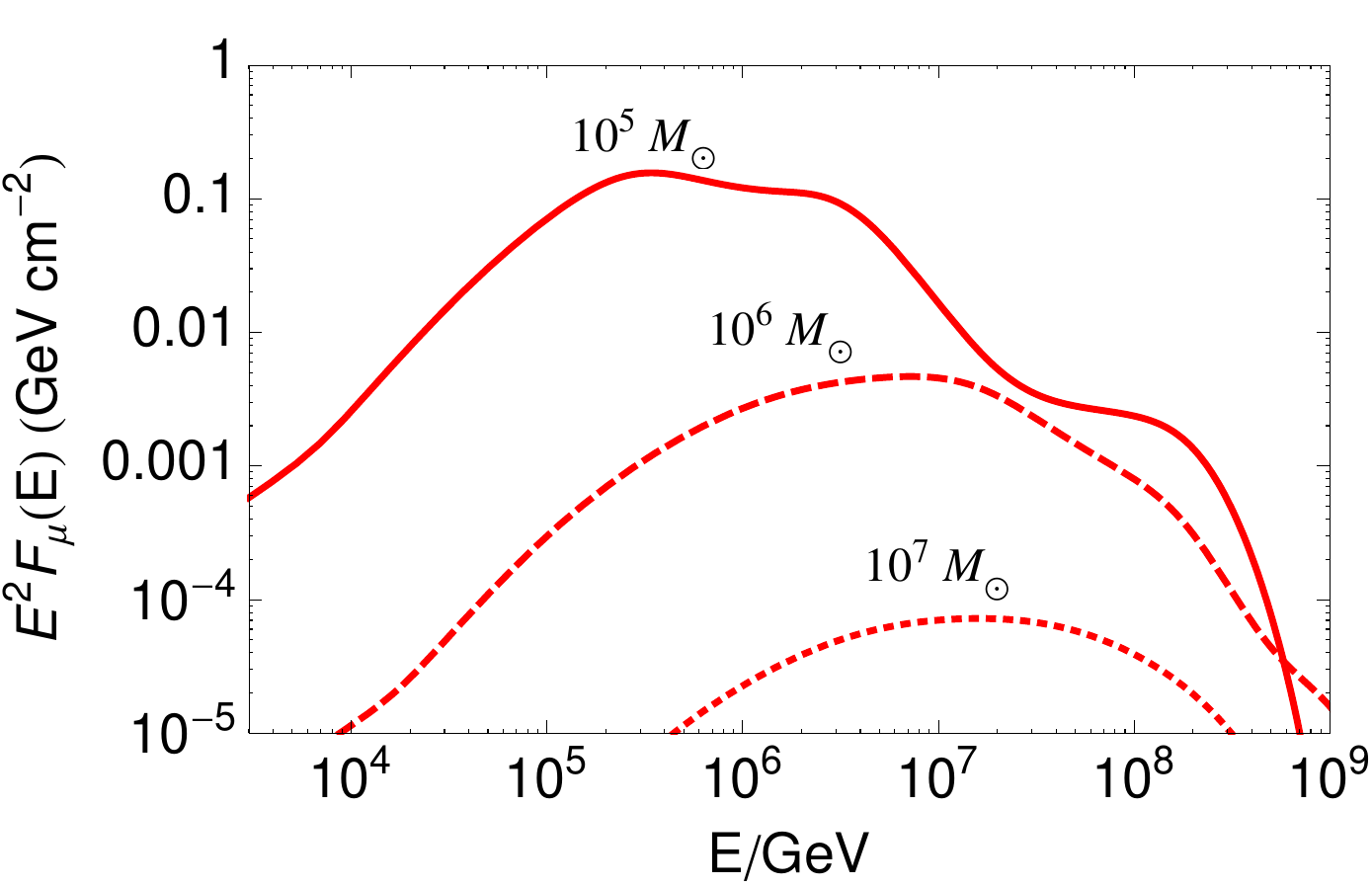}
\includegraphics[width=0.42\textwidth]{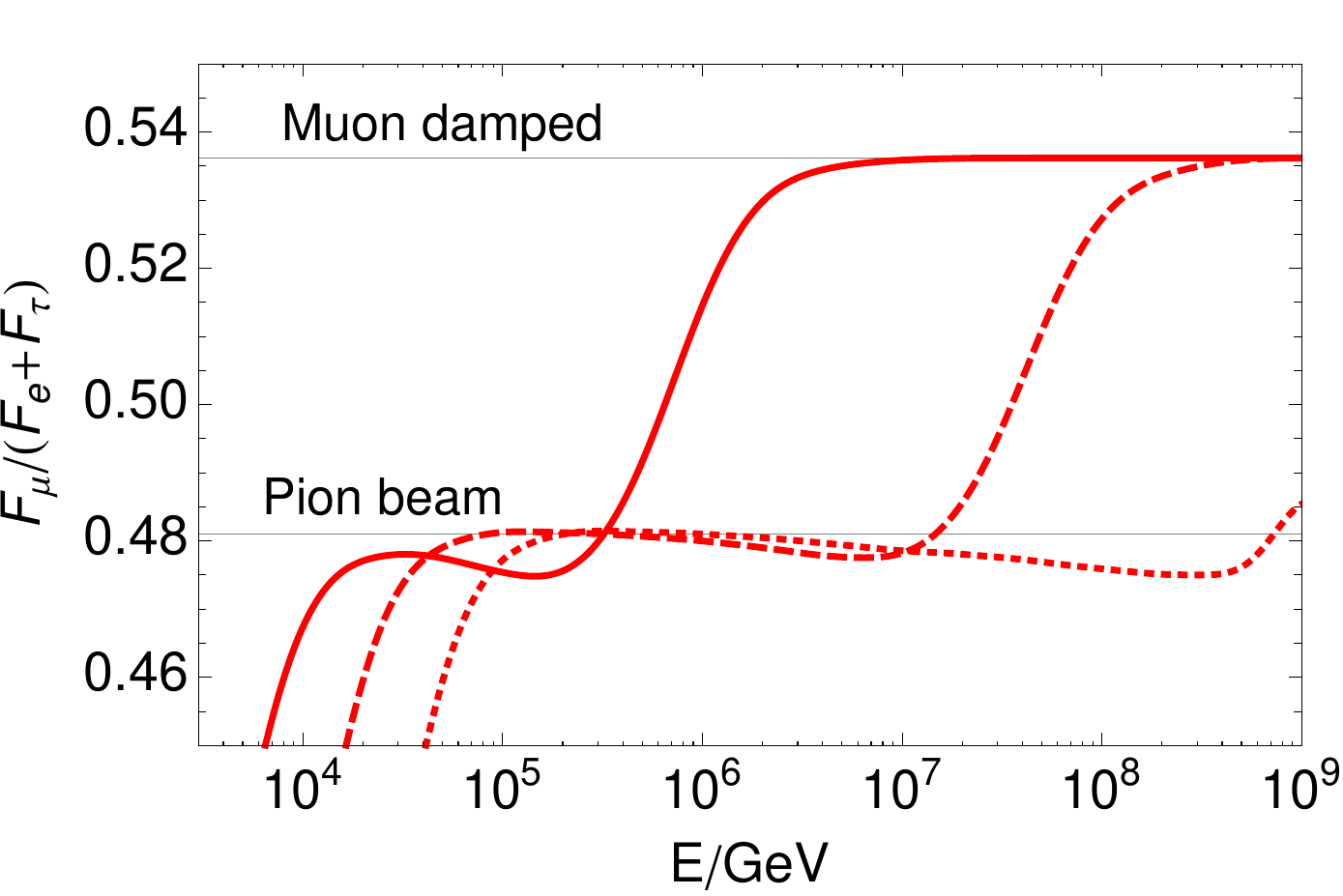}
\includegraphics[width=0.42\textwidth]{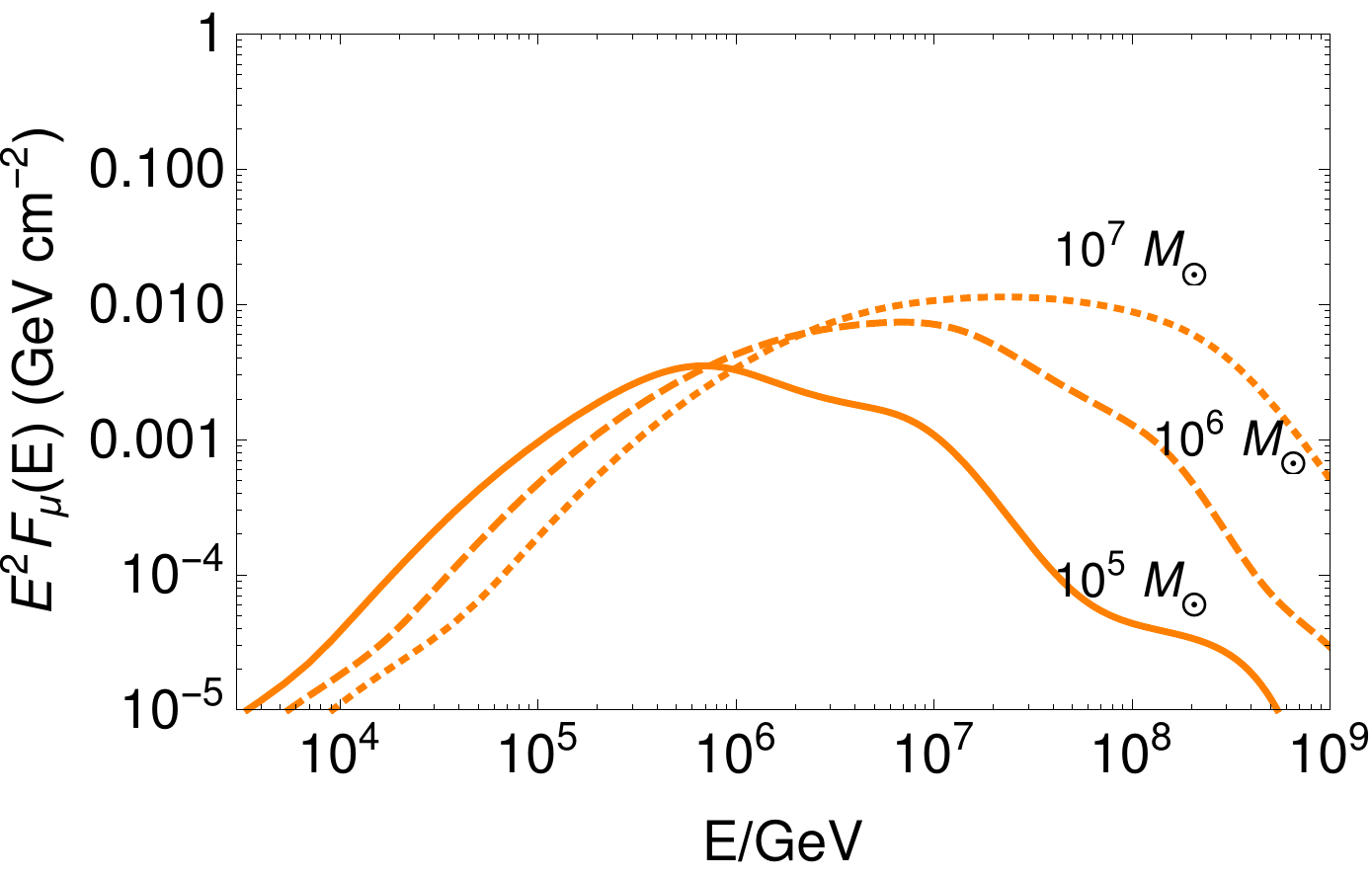}
\includegraphics[width=0.42\textwidth]{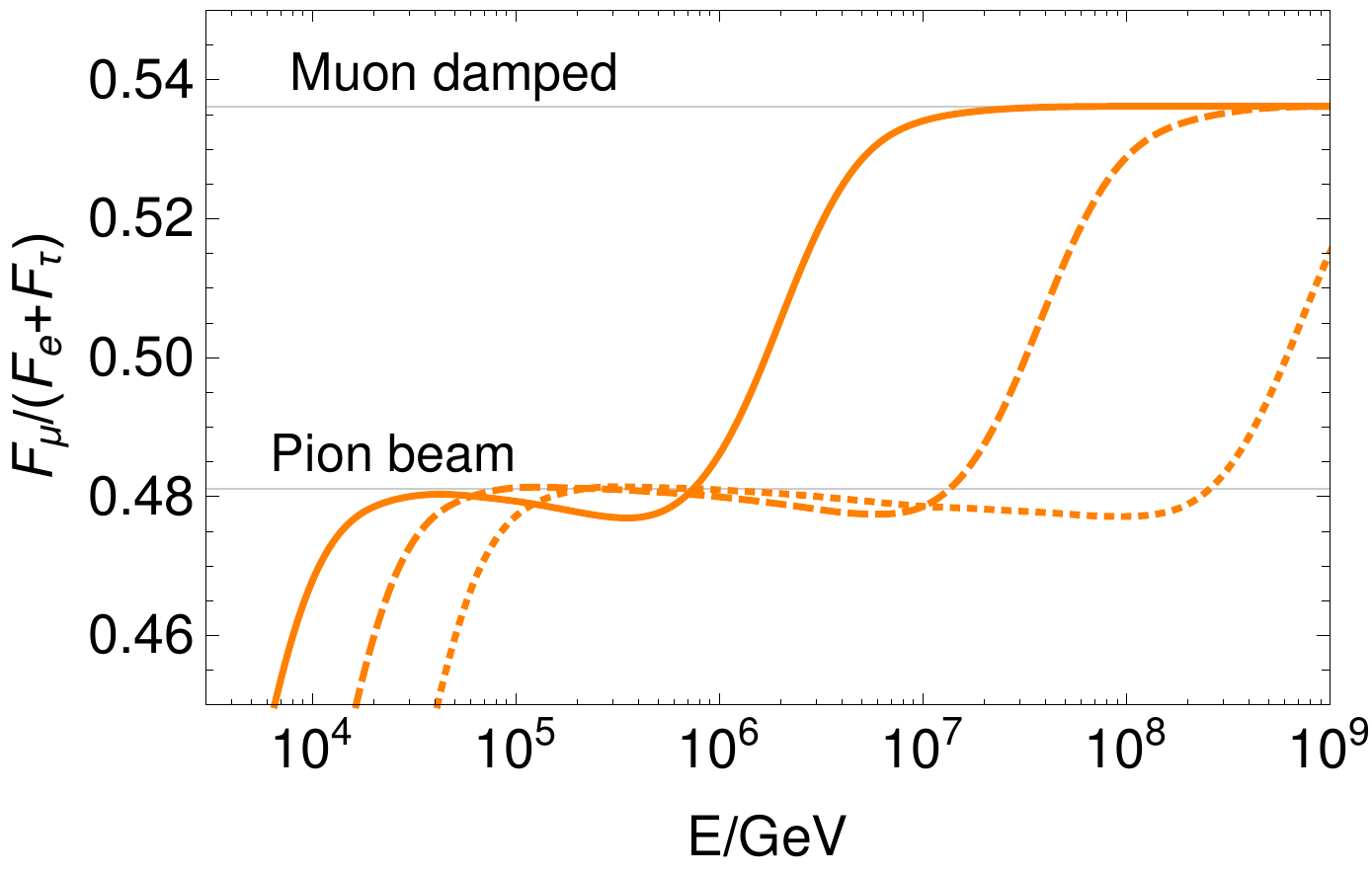}
\caption{The fluence of $\numu + \barnumu$ (left panels) and the flavor ratio (right panels), including flavor mixing, as a function of the \n\ energy, for a single \td\ at $z=0.35$. Shown are results for the Base, Weak, Strong, and Lumi cases (from top to bottom panes). In each figure, the curves  correspond to different \bh\ mass, $M=10^5  \msun$ (solid), $M = 10^6  \msun$ (dashed), $M = 10^7  \msun$ (dotted).  For the flavor ratio, the horizontal lines show the values expected for the standard pre-oscillation compositions $(F^0_e:F^0_\mu:F^0_\tau)=(1,2,0)$ (pion beam) and $(F^0_e:F^0_\mu:F^0_\tau)=(0,1,0)$ (muon damped source). 
} \label{fig:plotmasses}
\end{figure}

\figu{plotmasses} (left column) shows the fluence, $E^2 F_\mu$, of $\numu + \barnumu$ (after \oss) for a single \td, for the Base case and for the scaling scenarios outlined above and  for different values of $M$.  The  Base case result does not depend on $M$ because the neutrino flux is computed using $L_X$ and fixed values of $\Gamma$ and $t_\nu$.

The Weak case with $M=10^6 \msun$ differs from the Base one only for the value of $t_v$, which is one order of magnitude larger.  The \n\ fluence is suppressed by roughly the same factor, as expected from the suppression of the pion production efficiency (\Sec~\ref{subsub:scalings}, \App~\ref{app:analytical}).   From the scaling of $\Gamma$ (\eqs~(\ref{equ:gammafun}) and (\ref{equ:fpig})) a suppression by $\sim 10^{0.8}\sim 7$ is expected for every decade of increase of $M$; this matches the behavior in \figu{plotmasses} well at low energy.  The figure also shows the expected hardening of the \n\ spectrum with increasing $M$, due to the increase of the break energies (\Sec~\ref{subsub:scalings}).  

For the Strong case, the result for $M=10^6 \msun$ is very similar to the Base case, due to the nearly identical parameters.  As discussed in \Sec~\ref{subsub:scalings}, here the degree of enhancement with the decrease of $M$ is stronger; this means that low mass black holes will strongly dominate the neutrino flux because the time variability of the jet is assumed to be correlated with the period of the innermost stable orbit -- which is shorter for smaller \bh\ masses.
We also observe the expected stronger broadening of the spectrum towards higher energies due to the stronger scaling of the pion and muon break energies. 
In the Lumi scaling scale, the enhancement from small $M$ is compensated by the scaling $L_X \propto M$, which means that the flux actually increases with $M$ -- as discussed above.

In summary, for a single \td, in most cases, the quantity $E^2 F_\mu$ (with $F_\mu $ the fluence of muon \ns) peaks at $E \sim 10^6 -10^7$ GeV, with a maximum value $E^2 F_\mu \sim  10^{-5} - \mathrm{few} \times 10^{-3}~{\rm GeV cm^{-2}}$.     In the scenarios with parameter scaling, the \n\ flux increases with decreasing $M$.   In the case of Strong scaling, the fluence can be as high as $E^2 F_\mu \sim  10^{-1} ~{\rm GeV cm^{-2}}$ at $E \sim {\rm few \times 10^5} $ GeV.   

\figu{plotmasses} (right column) also illustrates the flavor ratio $f_\mu=F_\mu/(F_e + F_\tau) $ of  \n\ flux after flavor mixing in the three models considered here, and selected values of $M$.  This flavor ratio roughly corresponds to the observable ratio between muon tracks and cascades in IceCube, and has been widely used in the literature to study the impact of a change of the flavor composition.
For all models there is an energy window where $f_\mu \simeq 0.48$; this is the region where $\mu$ decay proceeds unimpeded, $E_\mu \lta E_{\mu,\mathrm{br}}$. In this regime, for each pion in the jet, two muon \ns\ and one electron \ns\ are expected, so that the original (before \oss) flavor composition of the \n\ flux is $(F^0_e:F^0_\mu:F^0_\tau)=(1:2:0)$.  At higher energy, $E_\mu \gta E_{\mu,\mathrm{br}}$, the ratio $f_\mu$ increases to $f_\mu \simeq 0.536$, reflecting the transition to the regime where $\mu$ absorption dominates over decay, so that the original flavor content is $(F^0_e:F^0_\mu:F^0_\tau)\simeq (0,1,0)$    While $E_{\mu,\mathrm{br}}$ is nearly the same for the Base and Weak cases, it is much lower for the Strong and Lumi scenarios at lower $M$, so that for $M =10^5\msun$, already at energies of a few PeV the \n\ flux enters the muon damped regime.

%%%%%%%%%%%%%%%%%%%%%%%%%%%%%%%%%%%%%%%%%%%
\section{Diffuse neutrino flux from \td} 
\label{sec:diffuse}

\subsection{Diffuse flux prediction: Spectrum and flavor composition}\
\label{sub:fluxearth}

The diffuse flux of \ns\ of a given flavor $\alpha$ from \tds\ -- differential in energy, time, area and solid angle --  is obtained by convolving the \n\ emission of a single \td\ with the cosmological rate of \tds\  (see, e.g. \cite{Ando:2004hc} for the formalism): 
\begin{equation}
\Phi_\alpha(E) =\frac{c}{4 \pi H_0} \int_{M_{\rm min}}^{M_{\rm max} } dM \frac{\eta}{2 \Gamma^2(M)} \int_{0}^{z_{\rm max}}  dz\ \frac{ \dot{\rho}(z,M) Q_\alpha(E(1+z),M)}{\sqrt{\Omega_M (1+z)^3+\Omega_\Lambda}}~,
\label{equ:jtde}
\end{equation}
where $Q_\alpha$ is the number of neutrinos emitted per unit energy in the \bh\ frame, and $E^\prime=E(1+z)$ is the \n\ energy in the same frame; $E$ is the energy observed at Earth. Here $\eta$ is the fraction of \tds\ that generate relativistic jets, which is assumed to be a constant, $\eta \simeq 0.1$. This value has been suggested as plausible on the basis of a possible similarity with AGN \cite{Burrows:2011dn}.     The beaming factor $ 1/(2 \Gamma^2(M))$ accounts for the fraction of jets along our line of sight. Its use in \equ{jtde}  is consistent with the fact that the same equation contains the physical comoving rate of \tds\ (not corrected for beaming or observational biases).
 Due to the decline of $ \dot{\rho}(z,M)$ with $z$ (\figu{tderates}), the flux $\Phi_\alpha$ depends only weakly on $z_{\mathrm{max}}$; here we take $z_{\mathrm{max}}=6$, which is the maximum value considered in \bh\ mass function calculation of Shankar et al. \cite{Shankar:2007zg}.

\begin{figure}[tbp]
\centering
\includegraphics[width=0.42\textwidth]{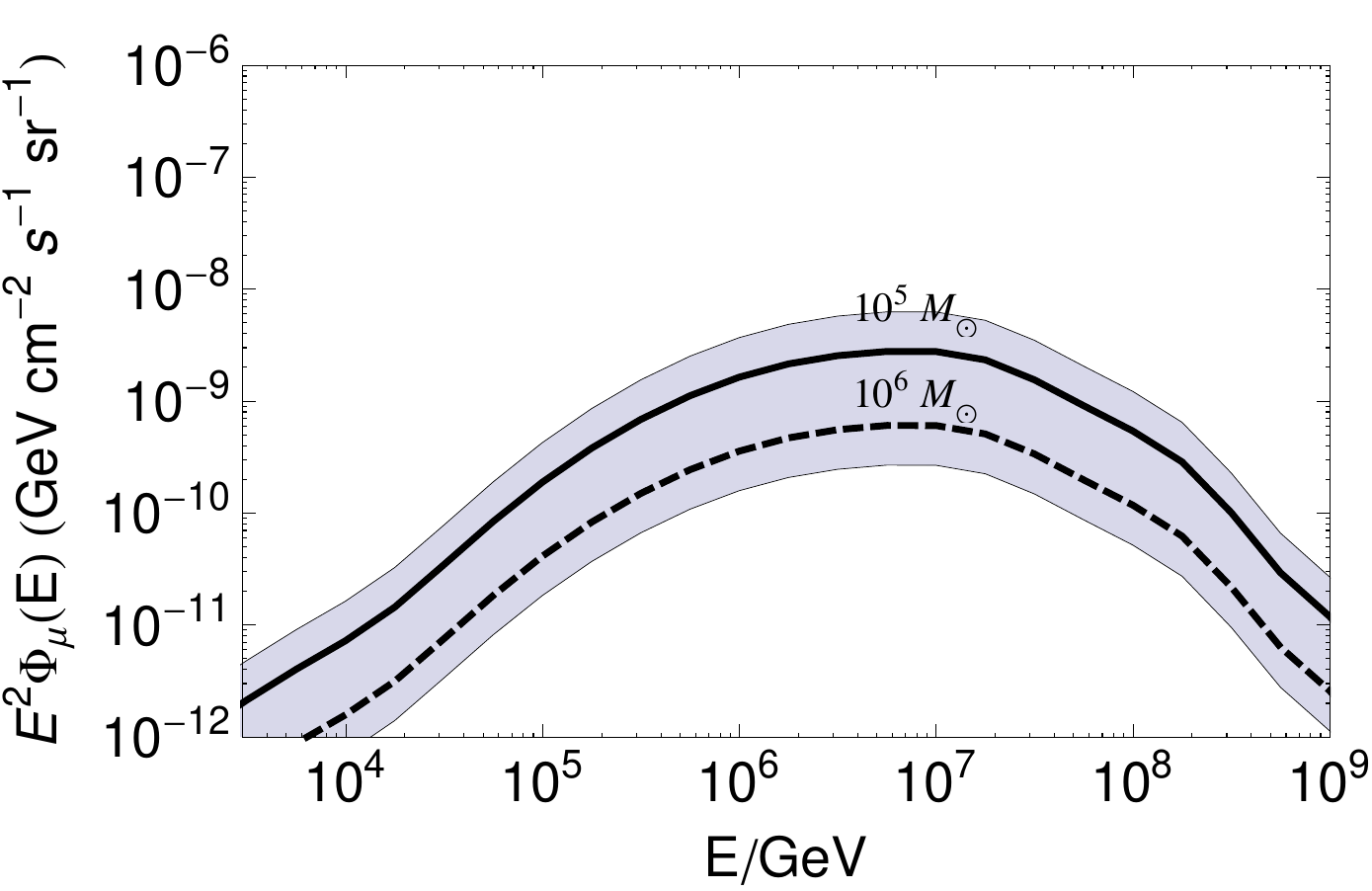}
\includegraphics[width=0.42\textwidth]{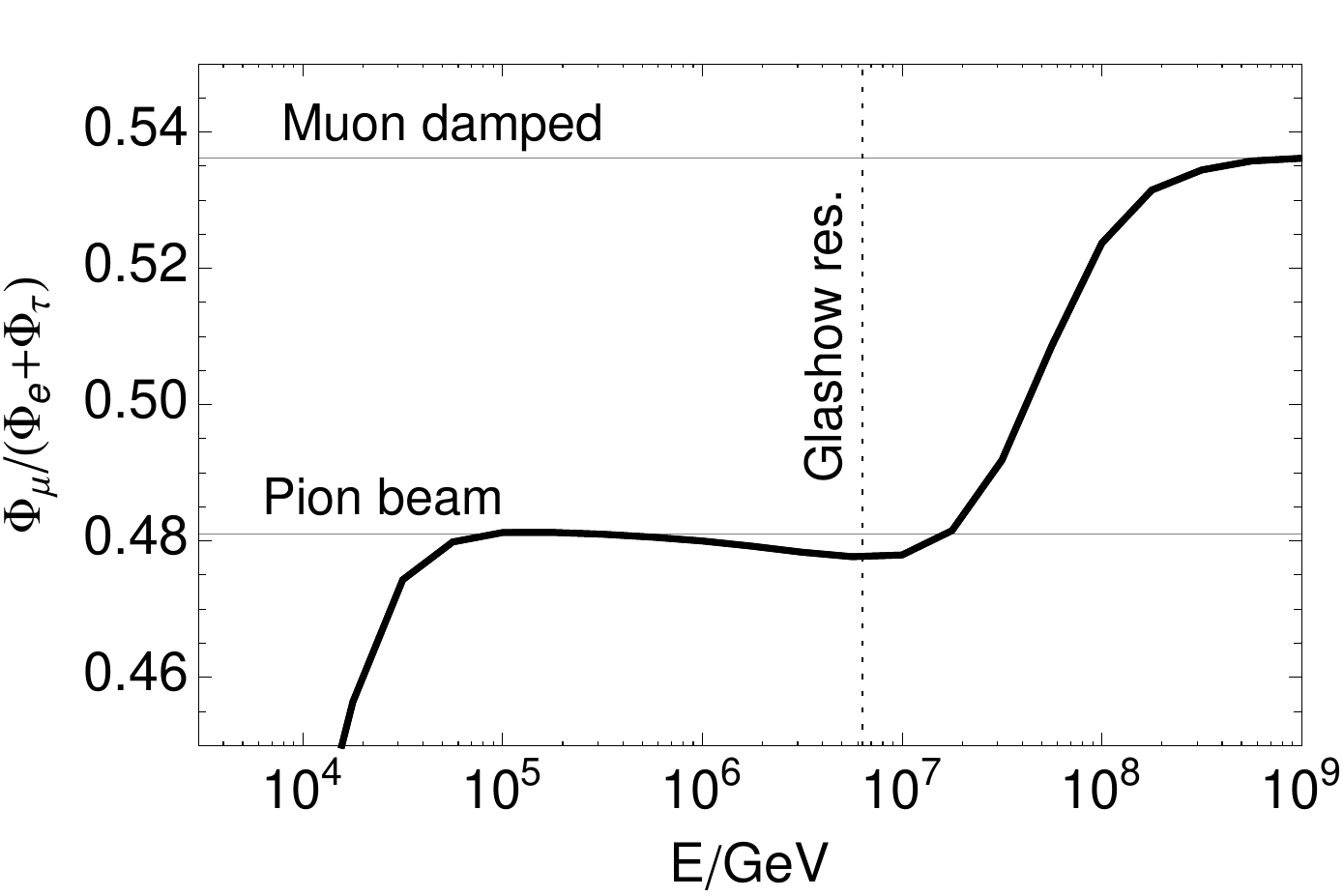}
\includegraphics[width=0.42\textwidth]{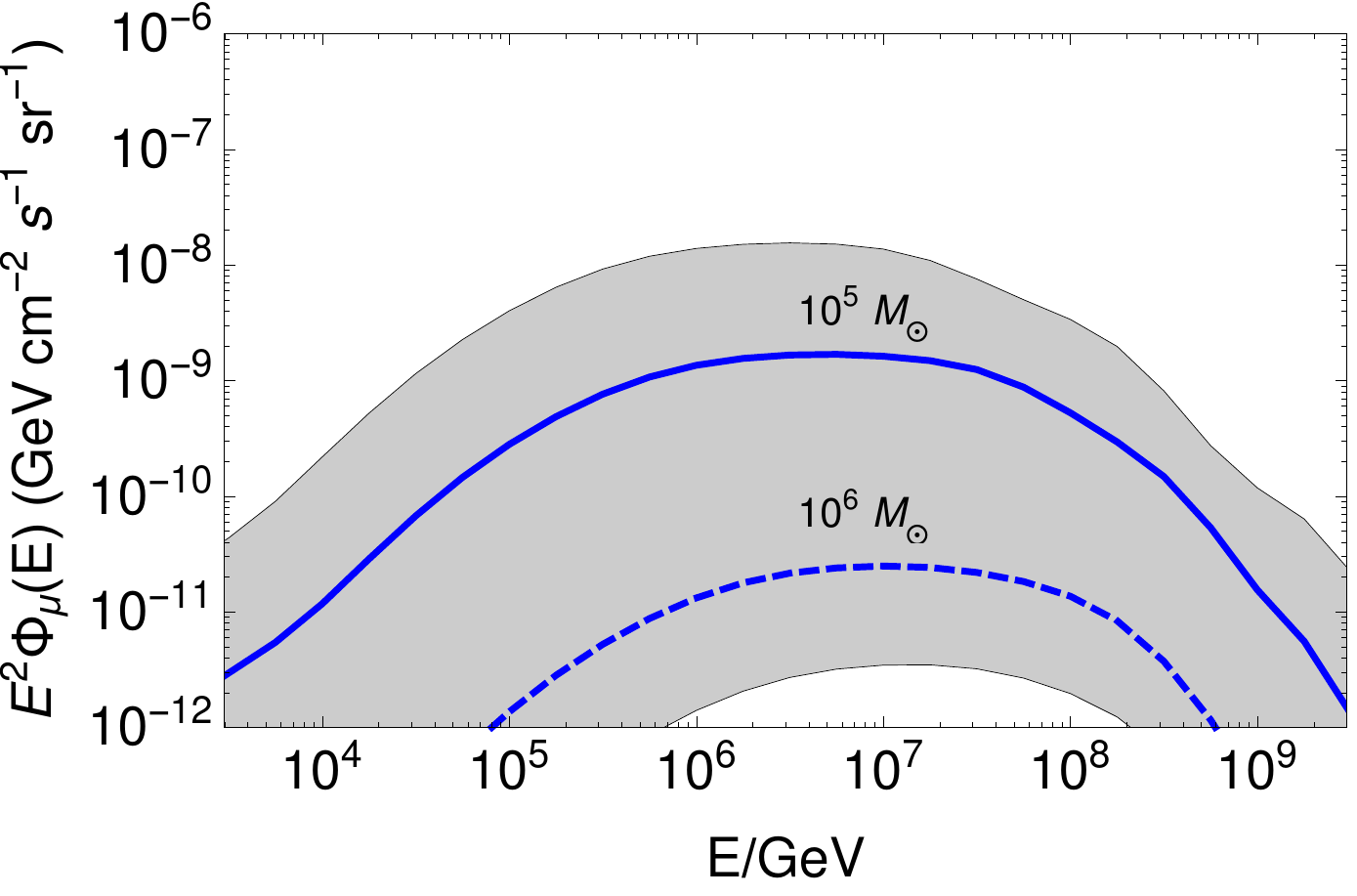}
\includegraphics[width=0.42\textwidth]{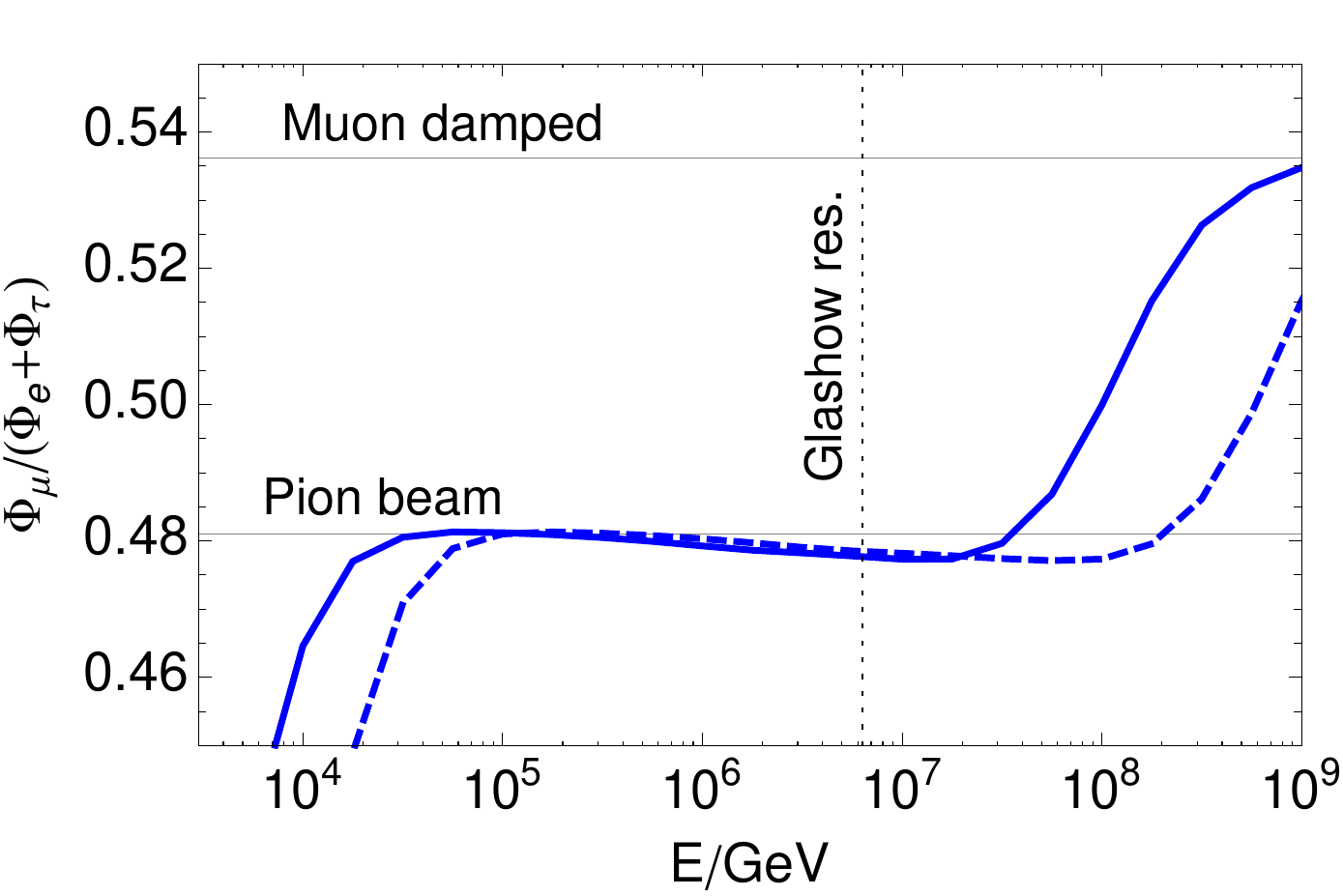}
\includegraphics[width=0.42\textwidth]{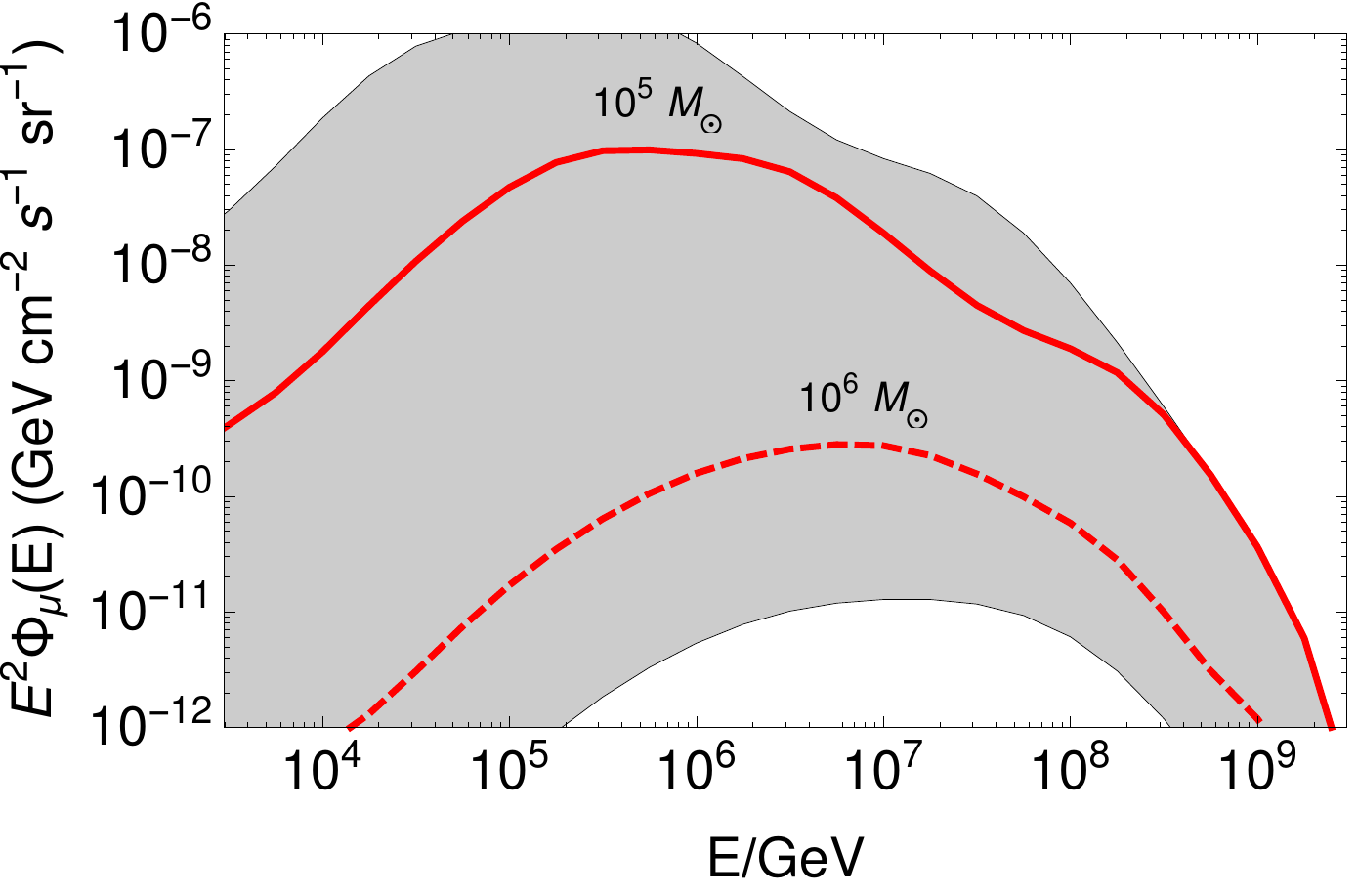}
\includegraphics[width=0.42\textwidth]{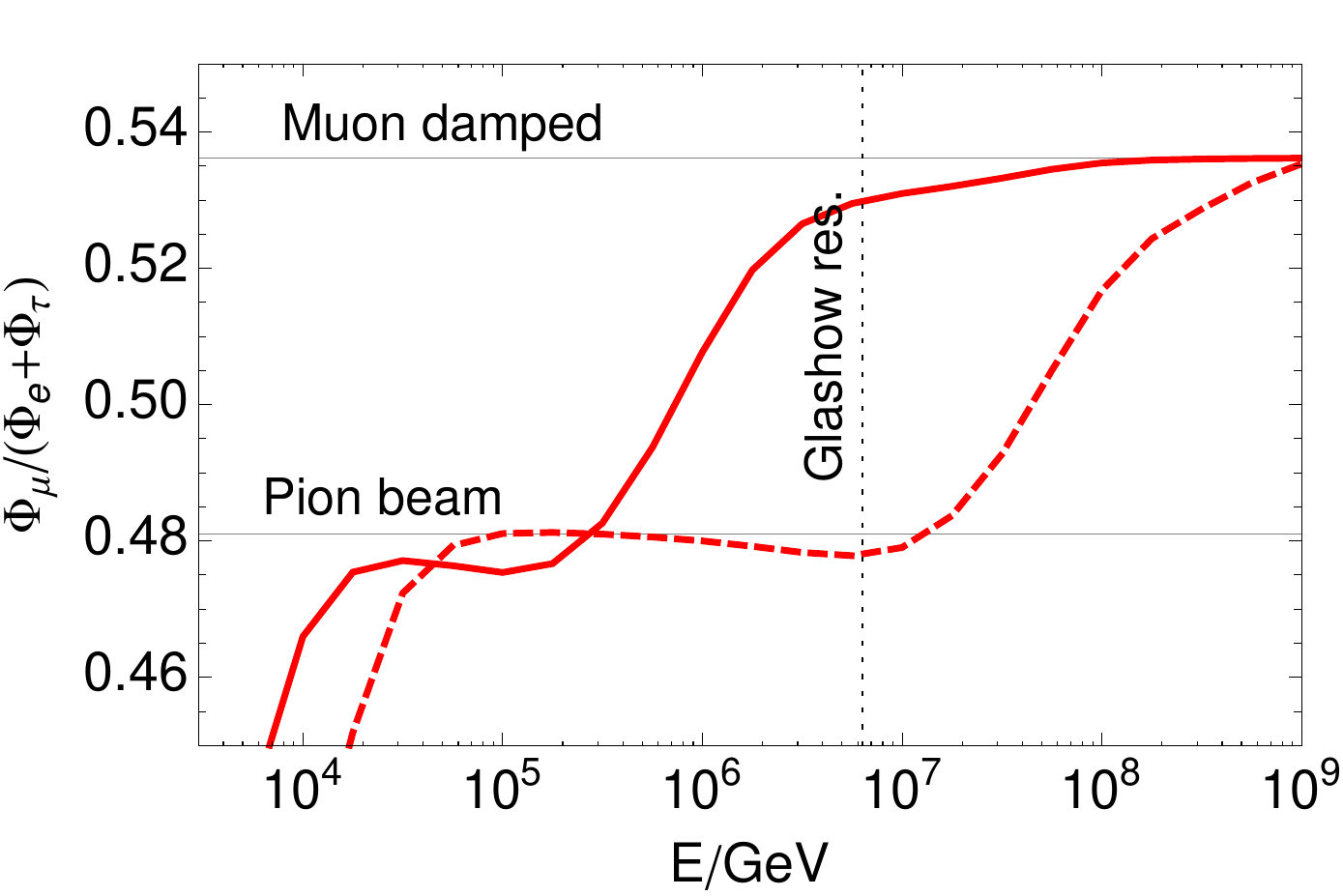}
\includegraphics[width=0.42\textwidth]{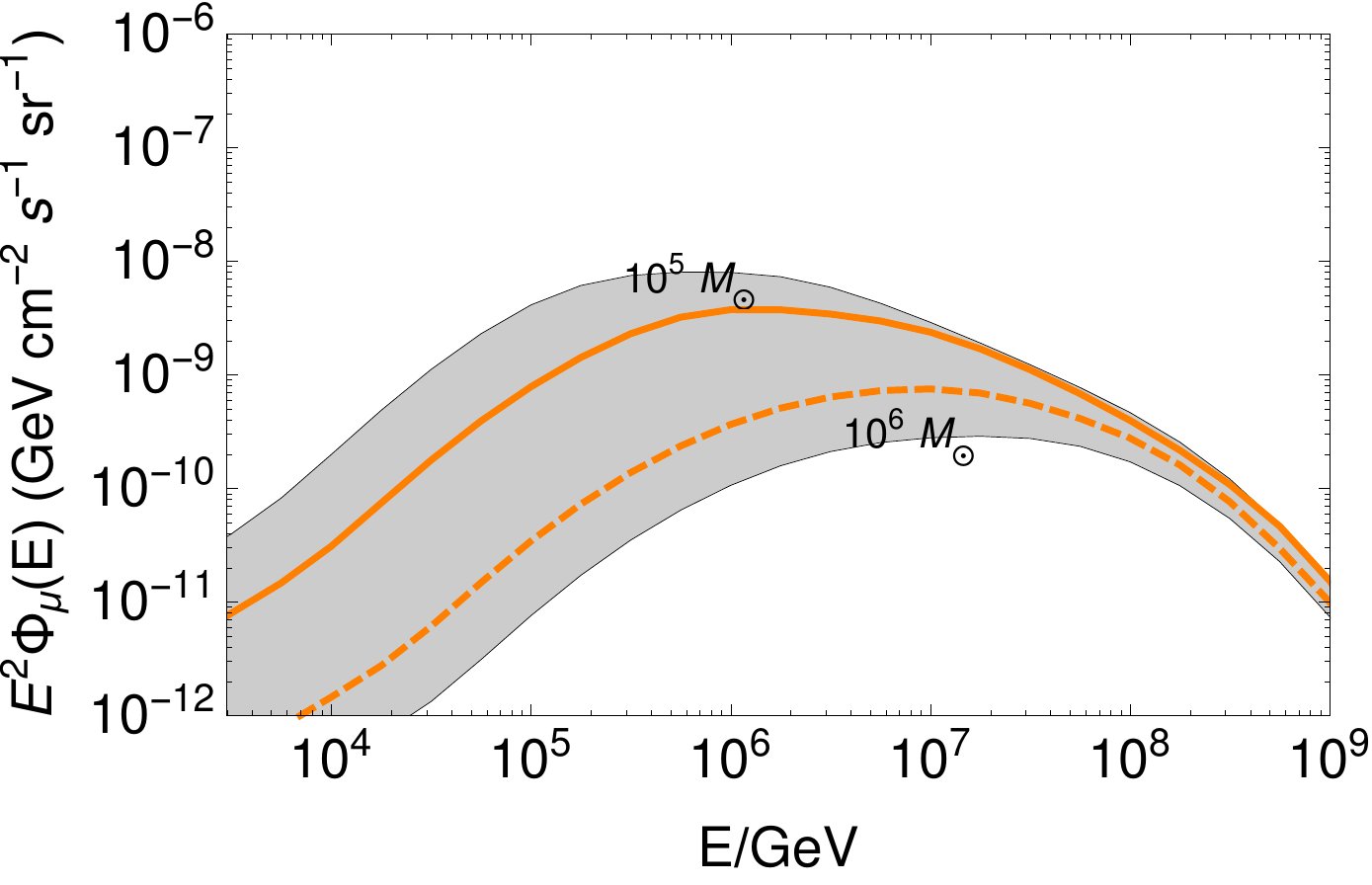}
\includegraphics[width=0.42\textwidth]{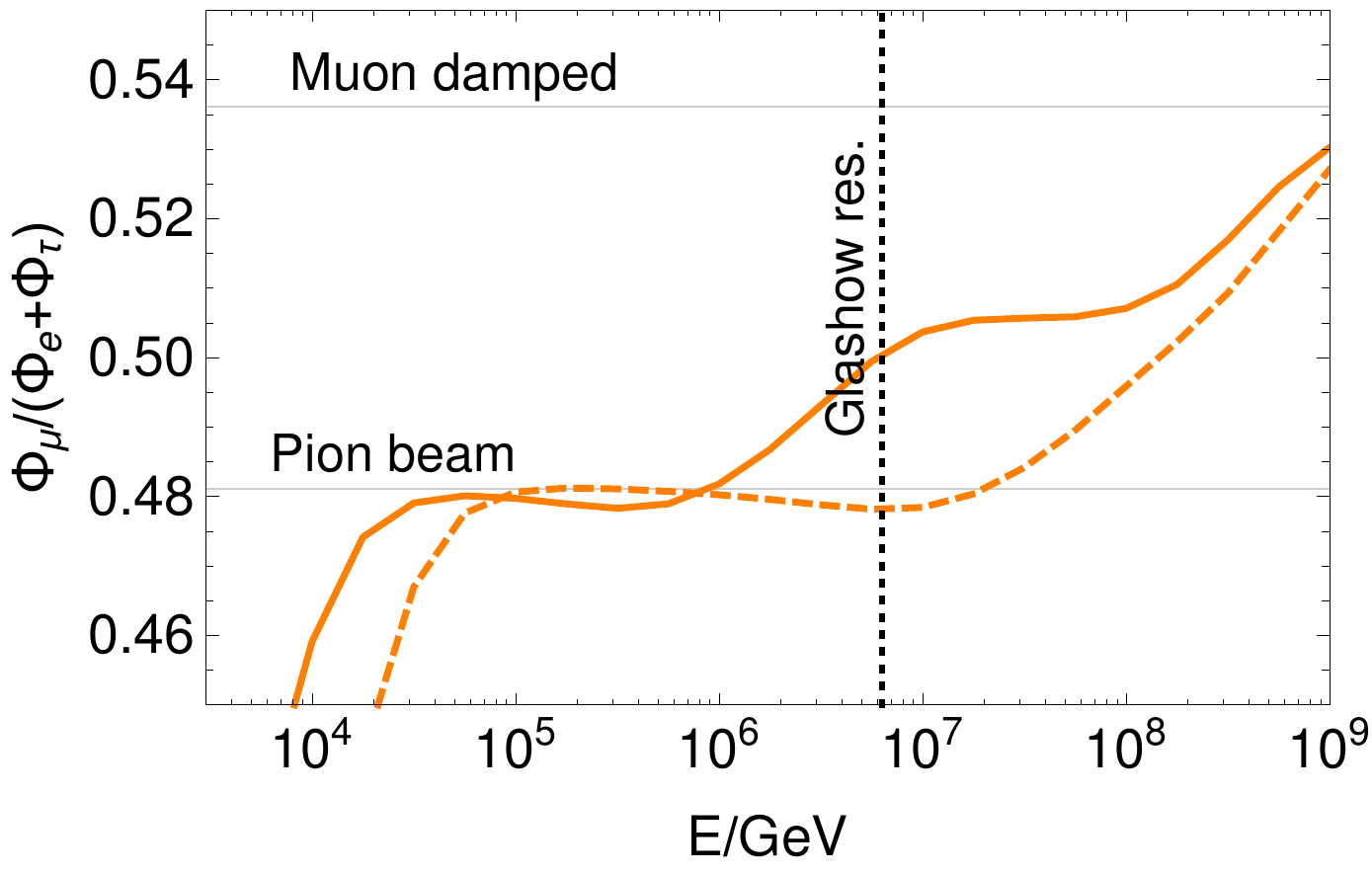}
\caption{The diffuse flux of $\numu + \barnumu$ (left panes) and the corresponding flavor ratio (right panes) at Earth, including flavor mixing, as a function of the \n\ energy, for the Base, Weak, Strong, and Lumi scaling cases (top to bottom), for $\mmin=10^{5}\msun$ (solid), and $\mmin=10^{6} \msun$ (dashed).  In the flux plots, the shaded regions show the variation corresponding to varying $\mmin$ in the interval $\mmin=[ 10^{4.5} ,10^{6.5} ] \msun$.  
In the flavor ratio figures, the horizontal lines show the values expected for the standard pre-oscillation compositions $(\phi^0_e:\phi^0_\mu:\phi^0_\tau)=(1,2,0)$ (pion beam) and $(\phi^0_e:\phi^0_\mu:\phi^0_\tau)=(0,1,0)$ (muon damped source).  The energy of the Glashow resonance in the flavor composition panels is marked by a dotted line.}
\label{diffuse}
\end{figure}

Fig. \ref{diffuse} shows the diffuse muon \n\ flux, $E^2 \Phi_\mu(E)$, for the four scaling scenarios of interest, and $\mmin=10^5,10^6\msun$ (solid curves).  The shaded area models the uncertainty on $\mmin$, which is varied in the interval   $\mmin=[ 10^{4.5} ,10^{6.5} ] \msun$.   In all cases, the spectrum resembles the spectrum of a single \td\ with $M \sim \mmin$ and $z\ll 1$ (\figu{plotmasses}), as expected since the \td\ rate is a decreasing function of $z$ and $M$ (\figu{tderates}). For the Base, Weak and Lumi cases (with $\mmin=10^5\msun$), the diffuse flux has a maximum of $E^2 \Phi_\mu(E) \sim 10^{-9}~{\rm GeV cm^{-2} s^{-1} sr^{-1}}$ between $1$ PeV and $10$ PeV.  For the Weak case, we note the stronger contribution of the lowest mass \bh, $M=10^5 - 10^6~\msun$, reflecting the more powerful \n\ emission as $M$ decreases (\Sec\ \ref{sub:flux}).  The same features are observed for the Strong case, with an even more enhanced contribution of the lowest mass \bh, which causes the flux to peak at lower energy, $E \sim 0.3 $ PeV. The dependence on $\mmin$ is, on the other hand, relatively mild for the Base and Lumi cases.

The post-oscillation flavor ratio for the diffuse flux is shown in Fig. \ref{diffuse} (right column). Like the fluence, it mainly follows the corresponding quantity for a single \td\ with lowest $M$ and lowest $z$ (\figu{plotmasses}). In the Lumi case, the contributions from lower and higher $M$ are comparable at about 10~PeV.
The observational implications of the energy dependence of the flavor composition will be discussed in the next section.

 Before closing this section, let us briefly comment on constraints on \tds\ from X-ray surveys. As a consistency check, in \App~\ref{app:xrayflux} we present the diffuse X-ray flux corresponding to the four scaling scenarios 
in Table~\ref{tab:cases}. This flux is found to be consistent with observations (see Appendix).

\subsection{Comparison to \ic\ data}
\label{sub:detection}

Let us now discuss the impact of current and future \ic\ data on the search for \ns\ from \tds.  After about 6 years of data taking, \ic\ has established that the Earth receives a flux of astrophysical \ns\ which is diffuse in nature, in first approximation, and at the level of $E^2 \Phi \sim \text{few} \times 10^{-8}~{\rm GeV cm^{-2} s^{-1} sr^{-1}} $, at observed energies between $\sim 30$  TeV and $\sim$~PeV.  Even accounting for large and poorly known uncertainties -- which depend in part on the model of the candidate sources --  this measurement appears to be in tension with the most extreme flux predictions in \Fig~\ref{diffuse}. In particular, for the parameters of reference used in this work, the Strong scaling scenario with $\mmin \sim 10^{4.5} - 10^5 \msun$ should be already strongly disfavored by the \ic\ data,  whereas all the other cases are compatible with \ic\ observations.

\begin{figure}[tbp]
\centering
\includegraphics[width=0.6\textwidth]{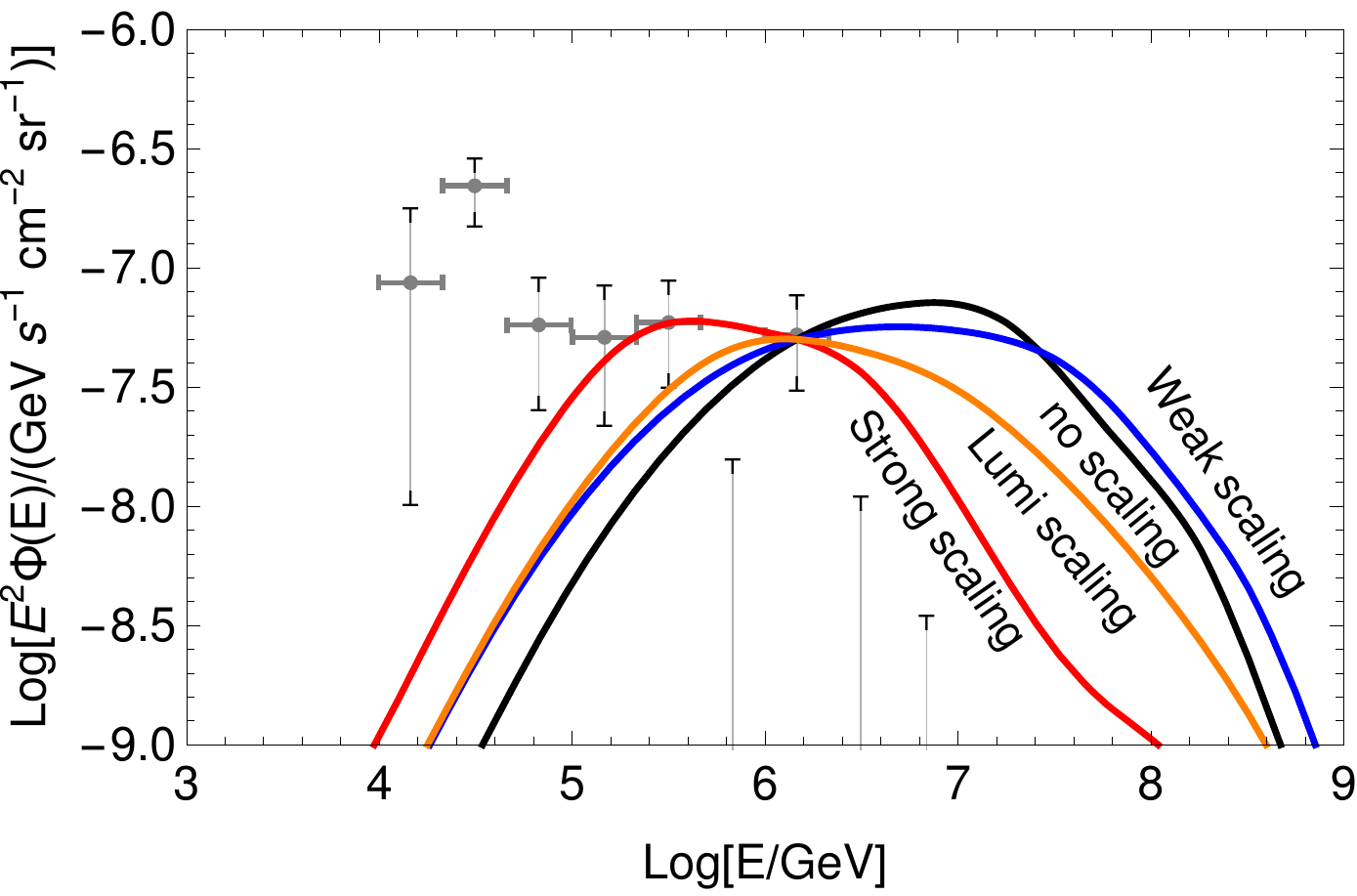}
\caption{The  spectra for the diffuse all-flavor flux, for the Strong, Weak, Lumi and Base (i.e., no scaling) models (labels on curves), for $\mmin=10^5 \msun$.   The overall constant $G=\xi_p \times \eta $ has been adjusted to saturate the measured \ic\ flux at $E\simeq {\rm PeV}$ (shown, data points \cite{Aartsen:2015zva,Aartsen:2015knd}), and takes the values $G=0.2, 10.9, 4.3, 8.1$ for the Strong, Weak, Lumi and Base cases, respectively.    }
\label{scaledplot}
\end{figure}

One should consider, however, that the jet parameters have a wide range of plausible values, which leads to a 
 a more quantitative question: can \tds\ account for most of the \ic\ flux, and for what values of the parameters? 
  From \eqs~(\ref{equ:protonorm}) and  (\ref{equ:jtde})  we see that $\Phi_\mu$ scales directly with 
\begin{equation}
 G \equiv \xi_p \times \eta \simeq 10 \times 0.1 \simeq 1 \, ,
\label{equ:gdef}
\end{equation}
evaluated for our standard assumptions. In principle, the neutrino flux also scales with the X-ray luminosity  (as both initial proton and the pion production efficiency are proportional to $L_X$), the beaming factor, the minimal and maximal proton energies (as the proton total energy is distributed over that energy range), \etc. However, these scaling factors are less trivial to treat, as, for instance, a higher luminosity will not only increase the pion production, but also the magnetic field and therefore the secondary cooling -- which partially compensates for that. We therefore do not include them directly in $G$.

Fig. \ref{scaledplot} shows the all-flavor fluence for the diffuse flux, for $\mmin=10^5 \msun$, with the factor $G$ adjusted (values in the figure caption) to saturate the \ic\ data (shown as well) at $E \simeq 1$ PeV. 
Note that we do not perform a statistical analysis of the data, but rather the 
normalization of the predicted flux is chosen so that the data point at 1~PeV is exactly reproduced.  We see that
the Base and Weak cases can describe the data in the 0.1-1 PeV energy range, although at the price of invoking parameter values $G \sim 8-11$. 
Such an increase of the neutrino flux could come from a factor of ten higher baryonic loading than anticipated in \Tab~\ref{default}, \ie, $\xi_p \sim 100$, or, equivalently, from a higher value of $\eta$.

Such a large baryonic loading may on the one hand not be unreasonable, as similar values are found for gamma-ray bursts fitting the UHECR data~\cite{Baerwald:2014zga}. 
In our notation, one can easily compare the energy in baryons with the constraint on the energy \equ{emax}: For the chosen $E_X$ and the conservative estimate $\Gamma \gtrsim 6$ (for $M> 10^5\msun$), one finds $\xi_p \lesssim  2 \Gamma^2 E_{\mathrm{max}}/E_X \sim  200$ in order to not to violate the constraint on the maximal emitted energy. This constraint is satisfied here, but the jets will have to be dominated by baryons for low mass black holes. However, increasing $\xi_p$ versus $\eta$ has the problem that it increases the tension with the multiplet constraints in IceCube, whereas increasing $\eta$ versus $\xi_p$ changes the fraction of jetted versus non-jetted \tds .
In addition to requiring a somewhat extreme value of $\xi_p$, the Base and Weak cases are in overall tension with the data due to their relatively hard spectrum, which overestimates the flux above PeV while underestimating it at lower energy. Our nominal assumptions may therefore be more plausible.
In different words, for the Base case, we find that $1/8 \simeq 12\%$ of the observed flux in IceCube can be described by \tds , whereas we find $1/11 \simeq 9\%$ for the Weak case at an energy of about 1~PeV.

Instead,  the Strong case describes the data best: it reproduces the observed energy spectrum well, with only some tension with the data in the second lowest energy bin, and only slightly overestimating the flux at the highest energy; see \fig~\ref{scaledplot}.  The normalization leads to $G \simeq 0.2$, \ie,  parameters even more conservative than the reference reference values used in \equ{gdef}. For example, one may choose $\xi_p = 2$ and $\eta=0.1$, or $\xi_p=10$ and $\eta=0.02$.  
Regarding the spectral properties, one should keep in mind that in this scenario the \n\ flux is vastly dominated by the lowest mass black holes (see \fig~\ref{diffuse}, lower left panel), 
 with a strong dependence on $\mmin$.   For smaller $\mmin$, $\mmin \sim 10^{4.5} \msun$, the description of  the data becomes better due to the softening of the \n\ spectrum, and the value of $G$ required to saturate the measured flux decreases further.  The opposite effect (worse description of data) is expected for larger $\mmin$. 

The Lumi scaling case is found to be in between these two scenarios. At the nominal prediction, 23\% of the observed IceCube flux can be described by \tds . The flux can, however, be saturated if the baryonic loading or $\eta$ are slightly adjusted, such as $\xi_p \simeq 40$ and $\eta=0.1$. The spectrum describes the IceCube data with slightly larger cutoff energy.
 
It is interesting to compare our prediction to current IceCube data fits. For a global analysis of data~\cite{Aartsen:2015knd}, relatively soft spectral indices of the neutrinos are found: $\alpha \simeq 2.5$ for a power law fit. A recent through-going muon analysis, however,  indicates a spectral index $\alpha \simeq 2$~\cite{Aartsen:2016xlq}. These findings, together with information on the spatial distribution of the events, suggest the possibility that at low energies a softer, possibly Galactic contribution dominates (\cf, low energy datapoints in \fig~\ref{scaledplot}, which cannot be reproduced), whereas at high energy, an extragalactic component dominates and the spectrum becomes harder~\cite{Palladino:2016zoe,Palladino:2016xsy}. The diffuse flux  from \tds\ is an example for such an extragalactic hard component.

It is especially noteworthy that the Strong and Lumi scaling cases have a unique signature, apart from the good   description of the spectral shape:  in this case the flavor composition changes from a pion beam to a muon damped source at $E \sim$ PeV  (see lower right panels of \fig~\ref{diffuse}). This indicates the transition to a regime, as the energy increases, where muons cool faster by synchrotron losses than they can decay, see \eq~(\ref{emubreak}). While this effect can probably not be seen in the current IceCube experiment, it might be visible at the planned volume upgrade IceCube-Gen2~\cite{Aartsen:2014njl}; see \Ref~\cite{Bustamante:2015waa} (Figs.~3 and 9 there). 

Perhaps even easier to test is the fact that the diffuse flux becomes muon damped at the Glashow resonance, see vertical lines in right panels of \fig~\ref{diffuse}. This issue is discussed in \Ref~\cite{Biehl:2016psj}: if the spectrum is hard enough, Glashow events must  be seen in the current IceCube experiment after about 10 years of operation even in the $p\gamma$ case under realistic assumptions for the photohadronic interactions. 
The most plausible scenario which can evade this constrained is a muon-damped source at the Glashow resonance at 6.3~PeV, for which the $\bar \nu_e$ at Earth can only come from oscillated $\bar \nu_\mu$ from the $\pi^-$ contamination at the source. A non-observation of Glashow event rates in IceCube may therefore be a smoking gun signature for a muon damped source at the Glashow energy, and therefore TDEs as dominant source class - or alternatives, such as low-luminosity gamma-ray bursts~\cite{Murase:2008mr},  microquasars~\cite{Reynoso:2008gs}, or AGN nuclei~\cite{Winter:2013cla}. IceCube-Gen2 can then be used for more detailed source diagnostics.

\section{Summary and conclusions} 
\label{sec:disc}

We have studied the production of high energy neutrinos in baryonic jets generated in the tidal disruption events (TDEs) of stars by supermassive black holes (\bh). Using the NeuCosmA numerical package, detailed results have been obtained for the fluence and flavor composition of a \n\ burst from an individual \td, and for the diffuse flux of \ns\ of each flavor from all cosmological \tds.  
Jet parameters motivated by observations have been used, and  variations of these parameters over the diverse population of parent \bh\ -- in the form of scalings with the \bh\ mass $M$ -- have been studied. Four scaling scenarios have been considered, ranging from no scaling at all (all \td\ being identical in the \bh\ frame) to a strong scaling, where the bulk Lorentz factor $\Gamma$ has been varied in a way motivated by AGN observations, and the variability time scale $t_v$ has been assumed to be correlated with the innermost stable orbital period of the \bh. We have also considered a possible luminosity distribution function, related to the \bh\ mass distribution.  The dependence on the occupation fraction of \bh\ -- in the form of the minimum mass $\mmin$ of \bh\ that can be found in the core of galaxies -- has been studied as well.  

In summary, we find that: 
\begin{itemize}

\item The largest contribution to the diffuse \n\ flux is expected from the \bh\  with lower mass located at low redshift $z \lta 1$.  This is because the rate of \tds\  decreases with $M$, and with $z$ as well. The dominance of low $M$ \tds\ is stronger in the scenarios with parameter scaling, as discussed above, and weakened if the luminosity scales with \bh\ mass.   In all cases, the spectral features and flavor composition of the diffuse flux generally reflect the quantities of the lowest mass \bh, and therefore are very sensitive to the cutoff of the \bh\ occupation fraction, $\mmin$. 

\item For the jet parameters of reference (\Tab~\ref{tab:parameters}), and in cases with weak or no scaling,  \tds\ can be responsible for $\sim$10\% of the observed \n\ flux at \ic\ at an energy of about 1~PeV.  Instead, for the same parameters,  strong scaling and $\mmin \lta 10^5 \msun$, the nominal \n\ flux would exceed the \ic\ measurement, which means that this extreme situation is already disfavored by current data, and IceCube constrains $\mmin$, the baryonic loading $\xi_p$, and the fraction $\eta$ of TDE producing jets. 

\item As a consequence, more moderate parameters can be chosen for the strong scaling case -- which can describe both normalization and spectral shape of the observed diffuse flux at the highest energies. Examples are $\eta = 0.1$ and $\xi_p=2$, or $\eta=0.02$ and $\xi_p=10$. Note that  more frequent \tds\ with lower baryonic loadings can release a possible tension with constraints from the non-observation of neutrino multiplets.  We also find that a second, possibly softer contribution to the flux of different origin (possibly Galactic) is needed to account for the lowest energy \n\ events. 

\item For the strong scaling case, which describes the spectral shape best, the flavor composition changes with increasing energy, and  approaches  a muon damped source at $E \gta $ PeV. This signature may be detectable in the next generation upgrade IceCube-Gen2. In addition, recall that so far \ic\ has not observed any events at the Glashow resonance.  If this became a statistically significant suppression in the future, it could be a smoking gun signature for TDEs as dominant source, because it is expected for a muon damped source. 

\item If the luminosity of jetted \tds\ scales with the \bh\ mass -- in addition to the scalings of the Lorentz factor and of the variability time scale -- an intermediate case is found which describes the spectrum very well, which exhibits a flavor composition change at the Glashow resonance, and which can describe about one fourth of the observed diffuse IceCube flux at its nominal prediction -- or saturate the diffuse flux with a slight increase of $\xi_p$ or $\eta$.  This case corresponds to a X-ray luminosity distribution function $\propto L^{-2}_X$ (\Sec~\ref{subsub:scalings}). 

\end{itemize}

 Overall, we find \tds\ to be an attractive possibility to explain, at least in part, the still elusive origin of the observed neutrino flux. Indeed, they naturally fit a hypothesis that has recently emerged from data analyses: that the \ic\ signal might be due to relatively frequent, transient photohadronic sources with photon counterparts at sub-MeV energies only.  Upcoming, higher statistics data at \ic\ and its future evolutions (such as \ic-Gen2) could substantiate the \td\ hypothesis in a number of ways.  One could be dedicated searches for time- and space- correlations of \n\ events with known \tds, possibly to be done in collaboration with astronomical surveys.  Another way is more detailed studies of the diffuse flux, that could show transitions, as the energy increases, in the spectral index and flavor composition of the flux, thus indicating the presence of a distinct component at $\sim$ PeV, of different origin than the lower energy events.   

Of course, one should consider the large uncertainties that affect the prediction of the \n\ flux from \tds. We  illustrated some of them, especially those due to the uncertain low mass cutoff (i.e., the lower mass end of the \bh\ occupation fraction), and those associated to  the scaling of time variability  or of the Lorentz factor of the jet  with the \bh\ mass, the baryonic loading of the jet, the fraction of TDEs producing jets, and the uncertain luminosity distribution.  These quantities are likely to become better known as more astronomical data are gathered on \tds.  

It is also fascinating that \n\ detectors themselves might contribute to our learning of the physics of tidal disruption.  Indeed, flux constraints from \n\ data could establish important upper limits on the energetics, baryon content and and frequency of \tds; these limits would be  complementary to astronomical observations, which are more strongly affected by absorption and limited sky coverage.   Alternatively, a discovery of \tds\ as \n\ sources would most likely give upper limits on the low black hole mass cutoff, and would distinguish among different scaling models for the jet parameters with the black hole mass.   By probing tidal disruption events, \n\ data would therefore contribute to answering unresolved questions on the fundamental physics of black holes, on the birth and evolution of supermassive black holes, and on the dynamics of galactic cores that are usually quiet and are only ``illuminated" occasionally by tidal disruption. 

{\bf Note added.} During completion of this study, \Refs~\cite{Dai:2016gtz,Senno:2016bso} have appeared. Their conclusions (about $5-10\%$ of the diffuse flux observed at IceCube could be consistent with the TDE hypothesis) is roughly consistent with our result for the Base case ($1/8.12 \simeq 12\%$). \Ref~\cite{Dai:2016gtz} compute event rates from individual TDEs and derive a constraint on the diffuse IceCube flux. \Ref~\cite{Senno:2016bso} also considers neutrinos from chocked jets, and conclude that the contribution must be sub-dominant. Note that their $\tilde \xi_{\mathrm{cr}}$ corresponds to our $\xi_p$. Compared to \Refs~\cite{Dai:2016gtz,Senno:2016bso}, our work includes a fully numerical computation of the diffuse flux including flavor effects, and the scaling assumptions with the \bh\ mass function are unique to our work.

\subsection*{Acknowledgments}

We thank M. Ahlers, L. Dai, A. Franckowiak, M. Kowalski, K. Murase, A. Stasik and N. L. Strotjohann for useful discussions. 
CL is grateful to the DESY Zeuthen laboratory for hospitality when this work was initiated. She acknowledges funding from 
Deutscher Akademischer Austausch Dienst (German Academic Exchange Service), the National Science Foundation grant number PHY-1205745, and the Department of Energy award DE- SC0015406.
WW acknowledges funding from the European Research Council (ERC) under the European Union’s Horizon 2020 research and innovation programme (Grant No. 646623).

%%%%%%%%%%%%%%%%%%%%%%%%%%%%%%

\bibliographystyle{apsrev}
\bibliography{drafttde}

%%%%%%%%%%%%%%%%%%%%%%%%%%%%%%

\begin{appendix}
\section{Comparison to an analytical computation}
\label{app:analytical}

Here we compare our numerical computation to the analytical approach given in \Ref~\cite{Wang:2011ip}. We assume the same parameters as in \Tab~\ref{default} and the main text, unless explicitly stated. The following relationships among the observables are assumed to hold at $z=0$, where the oberver's frame corresponds to the \bh\ frame.

It is reasonable to approximate the pion, muon and and neutrino energies as fixed fractions of the parent proton energy, $E_\pi \sim 0.2 E_p$, $E_\mu \sim 0.15 E_p$ and $E \sim 0.05 E_p$. 
The \n\ flavor fluences  (without flavor mixings) can be modeled analytically as~\cite{Wang:2011ip} 
\beq
E^2 F^0_\mu (E) &=& \frac{1}{32 \pi d_L^2} \frac{E_X \xi_p}{\ln{(E_{p,\mathrm{max}}/E_{p,\mathrm{min}}})} f_{p\gamma} \zeta_\pi (1+\zeta_\mu)~, \nonumber \\
E^2 F^0_e (E) &=& \frac{1}{32 \pi d_L^2} \frac{E_X \xi_p}{\ln{(E_{p,\mathrm{max}}/E_{p,\mathrm{min}}})} f_{p\gamma} \zeta_\pi \zeta_\mu ~.
\label{fluxsimple}
\eeq
The pion production efficiency  $f_{p\gamma}$ is the average fraction of energy deposited into pion production. It is, similar to gamma-ray bursts~\cite{Waxman:1997ti,Guetta:2003wi}, given by
\be
f_{p\gamma} \simeq 0.35 \left(\frac{L_X}{10^{47.5}~{\rm erg \, s^{-1}}} \right) \left(\frac{\Gamma}{10} \right)^{-4} \left(\frac{t_v}{10^2~{\rm s}} \right)^{-1}  \left(\frac{\epsilon_b}{{\rm KeV}} \right)^{-1} \times 
\begin{cases}
(E_p/E_{pb})^{\beta-1}~~  {\rm for }~~E_p < E_{p,\mathrm{br}} \\
(E_p/E_{pb})^{\alpha-1}  ~~ {\rm for }~~E_p \geq E_{p,\mathrm{br}} 
\end{cases}~. 
\label{equ:fpig}
\ee
Here $L_X=E_X/\Delta T$ is the average X-ray luminosity, and $E_{p,\mathrm{br}}$ is the proton energy leading to photo-pion production at the $\Delta$-resonance corresponding to the X-ray break energy, \ie,
\be
E_{p,\mathrm{br}} =   1.5 ~10^7{\rm  GeV} ~\left(\frac{\Gamma}{10}\right)^2 \left( \frac{1~{\rm KeV}}{\epsilon_{X,\mathrm{br}}} \right) \, .
\label{epsilonb}
\ee

In \eq~(\ref{fluxsimple}), the factors $\zeta_\pi$ and $ \zeta_\mu$ are suppression factors that account for pion and muon propagation (energy losses and decay).  The quantity $\zeta_\pi$ can be expressed in terms of a spectral break energy $E_{\pi,\mathrm{br}}$, beyond which the timescale of synchrotron losses becomes smaller than the pion lifetime: 
\beq
&&\left\{\begin{array}{lr}
       \zeta_\pi = 1  & \text{for } E_\pi \lta  E_{\pi,\mathrm{br}} \\
       \zeta_\pi \propto E^{-2}_\pi & \text{for }  E_\pi \gta  E_{\pi,\mathrm{br}}
        \end{array} \right.  \nonumber \\
        & \text{with} & E_{\pi,\mathrm{br}} \simeq 5.8\times 10^8~{\rm GeV} \left( \frac{L_X}{10^{47.5}~\mathrm{erg \, s^{-1}}}\right)^{-\frac{1}{2}}\left( \frac{\xi_B}{1}\right)^{-\frac{1}{2}}\left( \frac{\Gamma}{10}\right)^4\left( \frac{t_v}{10^2~s}\right) ~. 
        \label{zetapi}
\eeq
A similar expression holds for $\zeta_\mu$, for with 
\beq
E_{\mu,\mathrm{br}} \simeq 3.1\times 10^7~{\rm GeV} \left( \frac{L_X}{10^{47.5}~\mathrm{erg \, s^{-1}}}\right)^{-\frac{1}{2}}\left( \frac{\xi_B}{1}\right)^{-\frac{1}{2}}\left( \frac{\Gamma}{10}\right)^4\left( \frac{t_v}{10^2~s}\right) ~. 
        \label{emubreak}
\eeq

\begin{figure}[tbp]
\centering
\includegraphics[width=0.7\textwidth]{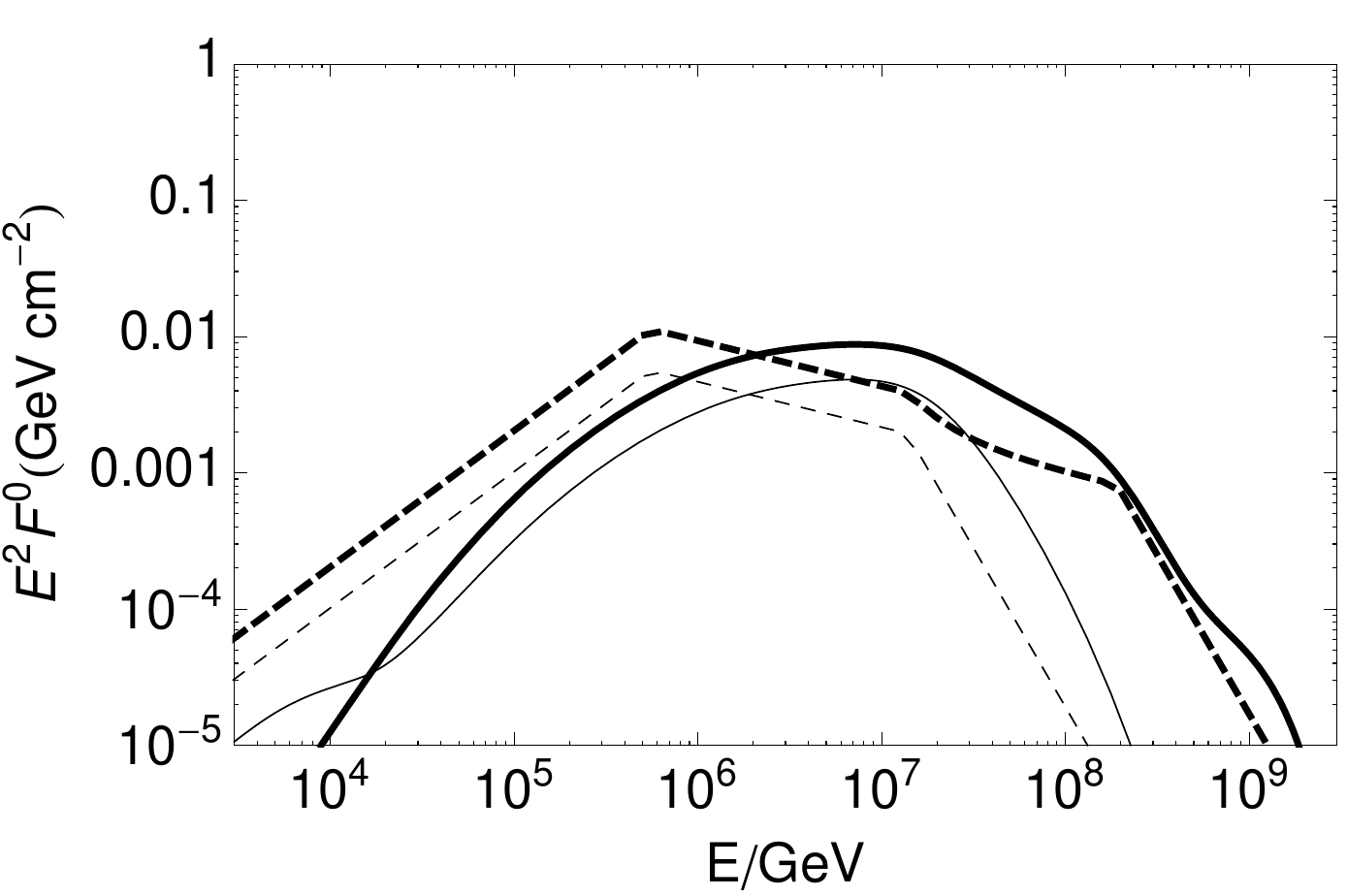}
\caption{The fluence $E^2 F^0_\alpha$ for $\numu + \barnumu$ ($\alpha=\mu$, thick) and $\nue + \barnue$ ($\alpha=e$, thin), for a \td\ at $z=0.35$.   Dashed: analytical approximation; solid: NeuCosmA numerical result. Flavor mixing is not included here, hence the factor of $\sim 2$ difference in the fluence compared to \figu{plotmasses}.
}
\label{fig:numericalvsanalytical}
\end{figure}

A comparison between the analytical technique and the numerical techniques  is given in \figu{numericalvsanalytical}. First of all, it is noteworthy that our numerical and analytical techniques match relatively well in terms of both shape and normalization. 

The slightly different shape comes mostly from high-energy photohadronic processes, see \Ref~\cite{Baerwald:2010fk}, from kaon production (at the highest energies), and from the different treatment of the photo-production threshold (at the breaks)~\cite{Hummer:2011ms}. At low neutrino energies, the analytical curves are simply extrapolated with \equ{fpig} using the high-energy photon spectral index $\beta$, whereas the numerical computation cuts off at a maximal photon energy given by the observed energy window (relevant for the minimal neutrino energy).  

The somewhat lower numerical versus analytical normalization is rather a coincidence of two competing processes: additional (to the $\Delta$-resonance) high-energy photomeson production modes enhance the pion production compared to the analytical estimate and make the spectral peaks more pronounced~\cite{Baerwald:2010fk}, whereas several reduction factors  have been identified in \Refs~\cite{Li:2011ah,Hummer:2011ms}. One example for the flux reduction is the over-estimation of the pion production efficiency using the break energy in \equ{fpig} instead of integrating over the whole spectrum. The relative normalization between analytical and numerical computation depends on the type of assumptions made for the analytical computation and the parameters (such as photon break energy and spectral indices). 

We observe that the analytical method matches the numerical computation relatively well (within about a factor of two), whereas for GRBs, the analytical computation typically overestimates the neutrino flux much more significantly.

%%%%%%%%%%%%%%%%%%%%%%%%%%%%%%%%%%%%%%%%%%%%%%%%%%%%%%%%%%%%%%%%%%%%%%%%%%%%%

\section{Predicted X-ray flux}
\label{app:xrayflux}

\begin{figure}[tbp]
\centering
\includegraphics[width=0.5\textwidth]{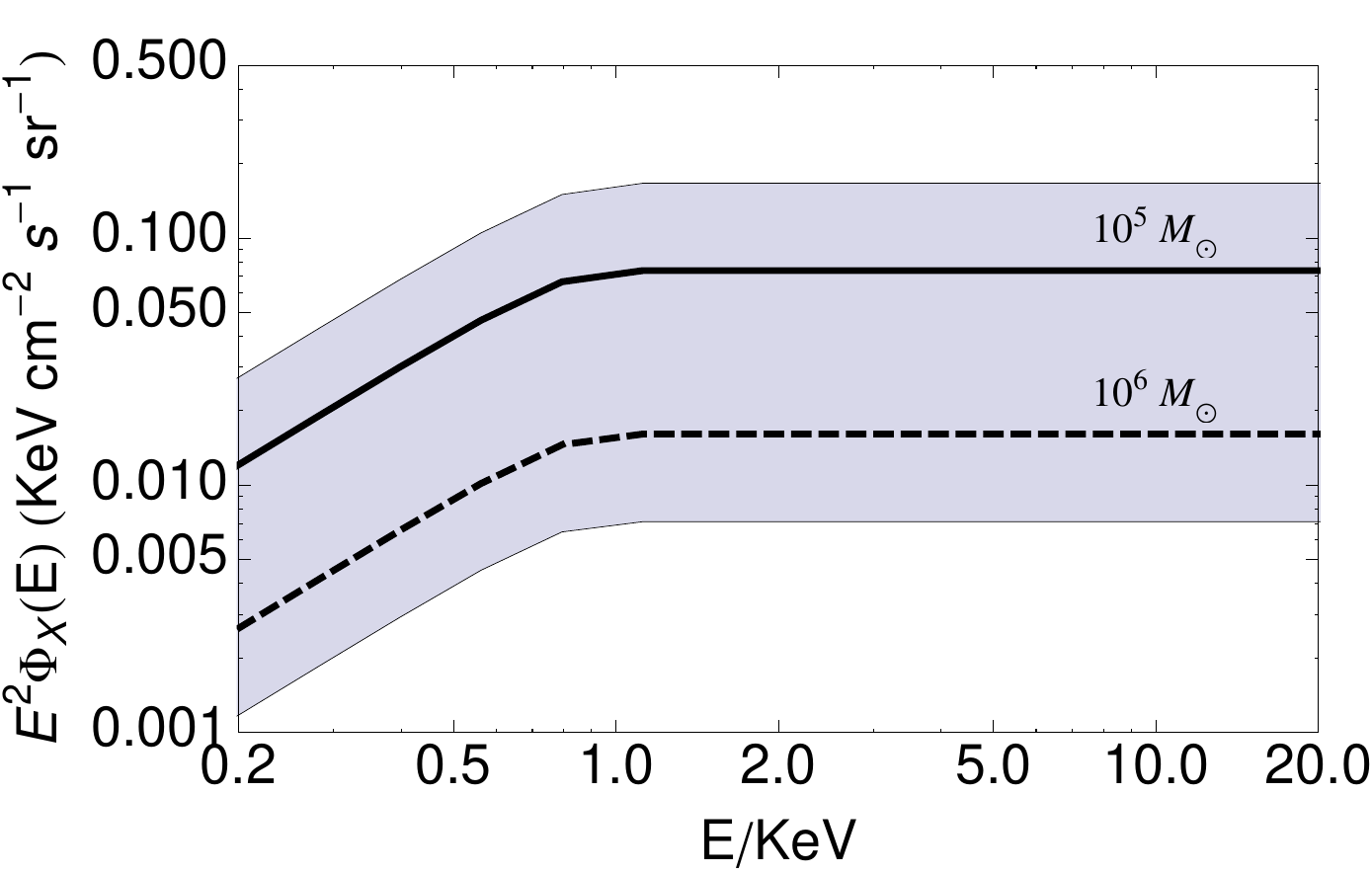}
\includegraphics[width=0.5\textwidth]{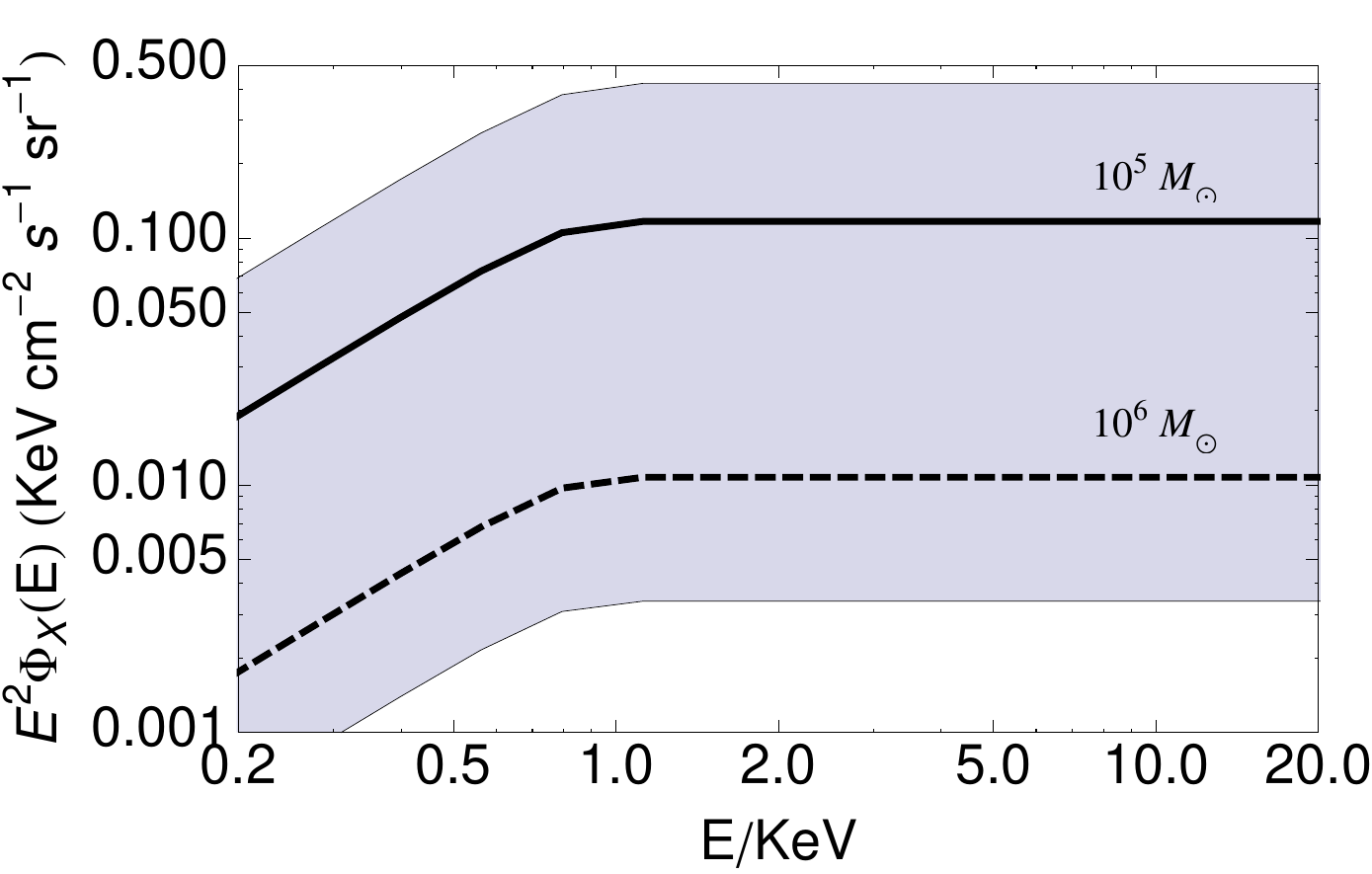}
\includegraphics[width=0.5\textwidth]{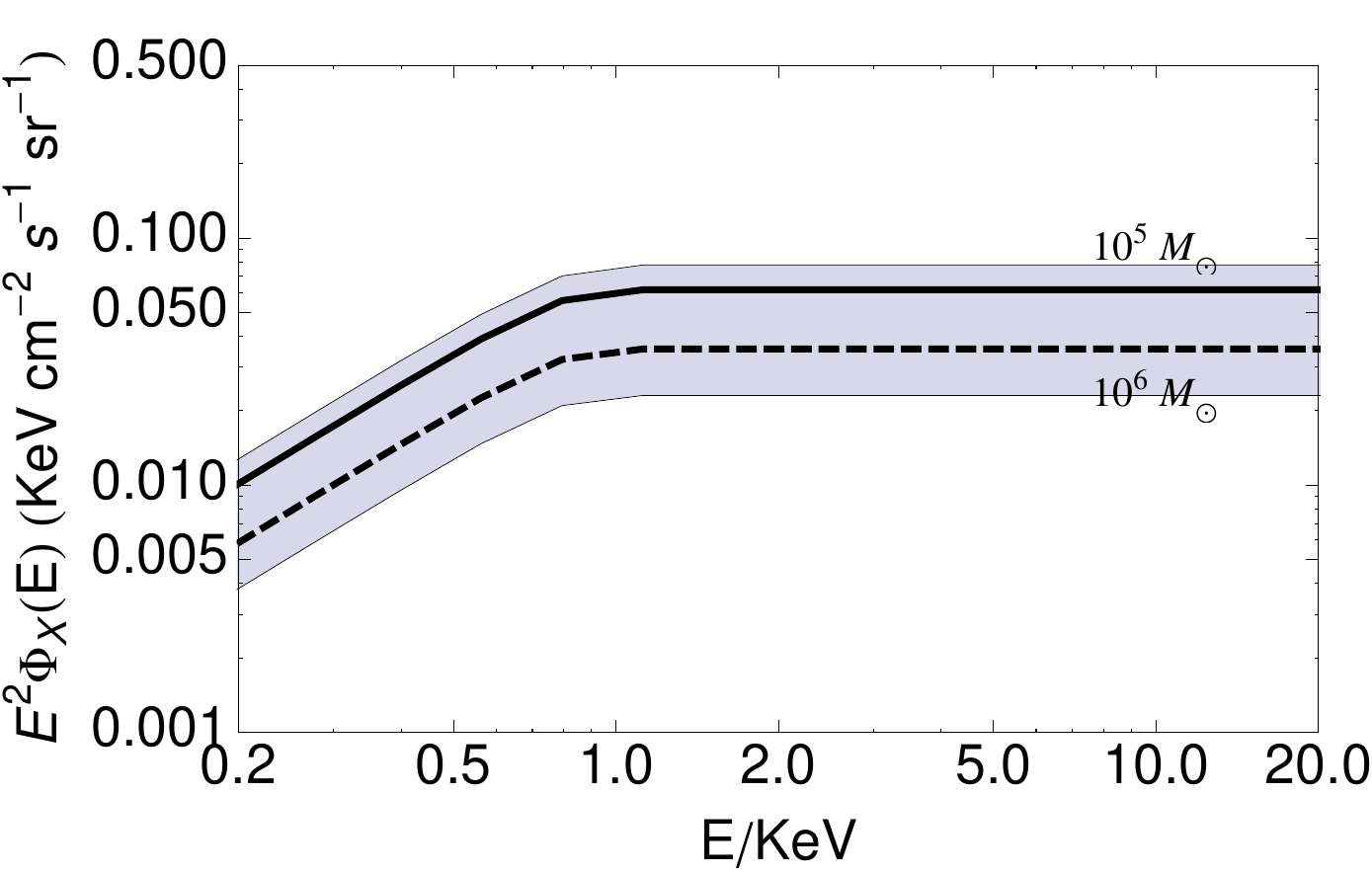}
\caption{The diffuse X-ray flux predicted in the Base case (upper pane), Weak and Strong case (middle pane) and Lumi case (lower pane), for $\mmin=10^{5}\msun$ (solid), and $\mmin=10^{6} \msun$ (dashed).  The shaded regions show the variation corresponding to varying $\mmin$ in the interval $\mmin=[ 10^{4.5} ,10^{6.5} ] \msun$. } 
\label{fig:xray}
\end{figure}

As a consistency check of our results, we calculated the X-ray flux expected for the different scaling scenarios (\Tab\ \ref{tab:cases}), following the same formalism as in \eq\ (\ref{equ:jtde}). 
 The results are shown in \fig\ \ref{fig:xray}. As expected from the scalings of $\Gamma$ and $L_X$, the contribution of the lower mass \bh\ is largest in the Weak and Strong cases, and suppressed in the Lumi case.  Note that the Weak and Strong cases differ only by $t_v$, which affects the neutrino production but not the X-ray flux. In all scenarios, the flux is consistent with observations, being at least one order of magnitude below the diffuse extragalactic soft X-ray flux as measured at $E \simeq 0.25$ KeV by ROSAT: $E^2 \Phi_X \simeq 5- 9~{\rm KeV~ cm^{-2} s^{-1} sr^{-1}}$ \cite{rosat}. At $E \gta {\rm KeV}$,  the observed diffuse flux is even larger (see e.g., \cite{Gilli:2003bm} and references therein), thus strongly outshining the predicted \td\ flux.    

\end{appendix}

%%%%%%%%%%%%%%%%%%%%%%%%%%%%%%%%%%%%%%%%%%%%%%%%%%%%%%%%%%%%%%%%%%%%%%%%%%%
\end{document}